\documentclass[aps,twocolumn]{revtex4}

\usepackage{graphicx,epsfig}% Include figure files
\usepackage{amsthm}
\usepackage{amsmath}
\usepackage{amssymb}
\usepackage{mathdots}
\usepackage[normalem]{ulem}

\newcommand*{\citen}[1]{%
  \begingroup
    \romannumeral-`\x % remove space at the beginning of \setcitestyle
    \setcitestyle{super}%
    \cite{#1}%
  \endgroup   
}

\begin{document}

\title{$\mathbb{Z}_4$ Parafermions in an interacting quantum spin Hall Josephson junction coupled to an impurity spin}
\smallskip
\author{Yuval Vinkler-Aviv, Piet W. Brouwer and Felix von Oppen}

\affiliation{\mbox{Dahlem Center for Complex Quantum Systems and Fachbereich Physik, Freie Universit{\"a}t Berlin, 14195 Berlin, Germany} }

\begin{abstract}
$\mathbb{Z}_4$ parafermions can be realized in a strongly interacting quantum spin Hall Josephson junction or in a spin Hall Josephson junction strongly coupled to an impurity spin. In this paper we study a system that has both features, but with weak (repulsive) interactions and a weakly coupled spin. We show that for a strongly anisotropic exchange interaction, at low temperatures the system enters a strong coupling limit in which it hosts two $\mathbb{Z}_4$ parafermions, characterizing a fourfold degeneracy of the ground state. We construct the parafermion operators explicitly, and show that they facilitate fractional $e/2$ charge tunneling across the junction. The dependence of the effective low-energy spectrum on the superconducting phase difference reveals an $8\pi$ periodicity of the supercurrent.
\end{abstract}

\maketitle

\section{Introduction}

The search for electronic systems that host exotic Majorana and parafermion quasiparticles has attracted significant research effort in recent years. These types of systems are characterized by a topologically protected ground-state degeneracy, that gives rise to non-Abelian exchange statistics between the quasiparticles~\cite{Leijnse_2012,Alicea_2012,Beenakker_2013}. The fractionalization of fundamental degrees of freedom into such exotic excitations and their nontrivial statistics is interesting both from a theoretical point of view and also as a potential basis for quantum computation~\cite{Kitaev_2003,Nayak_2008,Fendley_2012}.

Over the years, different proposals for realizing and observing this type of phenomenon in solid-state systems were laid out, among others surfaces of topological insulators~\cite{Fu_2008}, semiconductor wires~\cite{Oreg_2010,Lutchyn_2010}, edges of fractional quantum Hall systems or quantum spin-Hall systems~\cite{Lindner_2012,Clarke_2013}, and interacting nanowires~\cite{Sagi_2014,Klinovaja_2014}. One such setup is a Josephson junction comprised of the edge of a quantum spin-Hall insulator with time reversal symmetry breaking Zeeman splitting~\cite{Fu_2009}. The $2\pi$ periodicity of the Jospheson current with the phase across the junction is replaced by $4\pi$ periodicity, reflecting the existence of weakly coupled Majorana fermions that allow for single-electron transfer between the superconducting banks. This proposal, and the clear signature it promised, led to a focus of experimental efforts to observe $4\pi$ periodic Josephson currents, in what came to be known as fractional Josephson junctions~\cite{Wiedenmann_2016,Bocquillon_2016,Deacon_2017}.

The absence of time reversal symmetry in the junctions is central to the stability of the Majorana bound states and to the observation of the $4\pi$ periodicity. The physics of time reversal symmetric topological Josephson junctions was studied by Zhang and Kane~\cite{Zhang_2014} and by Orth {\it et al.}~\cite{Orth_2015} who observed that interactions between the electrons in the junction lead to $8\pi$ periodicity of the supercurrent as a function of the phase bias across the junction. Similar to the association of the $4\pi$ periodicity with Majorana fermions that allow tunneling of single electrons across the junction, these authors relate the $8\pi$ periodicity to $\mathbb{Z}_4$ parafermions that allow the tunneling of fractional $e/2$ charges. In this context, it is convenient to view the Majorana bound states as $\mathbb{Z}_2$ parafermions. However, in order for the low-energy physics to be adequately described by these parafermions, the setup considered by these authors requires very strong electron-electron interactions in the junction. The strong interactions in the junction favor the formation of magnetic order, and the domain wall between regions with magnetic order and regions with superconducting order carry the parafermions. This is similar to other suggestions for realization of $\mathbb{Z}_n$ parafermions, on the edges of fractional quantum Hall states that are coupled alternately to ferromagnets and superconductors~\cite{Lindner_2012,Clarke_2013}.

Recently, an alternative route to $8\pi$ periodicity in topological Josephson junctions was suggested by Peng {\it et. al.}~\cite{Peng_2016} and by Hui and Sau~\cite{Hui_2017}. These authors showed that a weak exchange interaction between the edge electrons and a quantum impurity spin, without breaking time reversal symmetry, leads to a similar $8\pi$ periodicity of the current. The question of strong exchange interactions was not directly addressed in these papers, which explicitly assumed weak coupling with respect to the superconducting gap. The analogy between the two models in the weak coupling regime naturally suggests that a similar resemblance holds in the strong coupling regime, and that the low-energy effective Hamiltonian in that case will give rise to stable $\mathbb{Z}_4$ parafermions.

Here we study time reversal symmetric junctions strongly coupled to an impurity spin as an experimentally accessible system for constructing and detecting $\mathbb{Z}_4$ parafermions. We show that even for weak electron-electron interactions, the coupling to the local impurity might lead to the formation of such parafermions, and study the different local coupling parameters that control whether they will appear. We continue by characterizing these parafermions and the four-fold degenerate ground-state subspace they span, as well as their signature on physical features such as the periodicity of the current as a function of the phase bias across the junction and charge tunneling.

The impurity may be realized in experiment using a quantum dot in proximity to the edge. Such devices would allow for relatively high control over microscopic parameters and are accessible to readout and measurement. As the parafermions are composed not only of the edge electrons but also of the impurity (dot) degrees of freedom, we believe that this feature may open experimental opportunities to detect and possibly manipulate these exotic particles.

The remainder of the paper is organized as follows. We begin, in Sec.~\ref{sec:review}, by reviewing the underlying physical picture and the main results of the paper. Then in Sec.~\ref{sec:Model_and_Bos} we introduce the setup and map its Hamiltonian onto a bosonized version with conjugate superconducting and magnetic order parameters. We apply a renormalization group analysis of the bosonic Hamiltonian in Sec.~\ref{sec:RG}, and study the different strong-coupling regimes that may describe the low-energy behavior of the system, depending on the microscopic parameters. We then proceed to study the strong-backscattering limit in Sec.~\ref{sec:Strong_J}. We establish the ground-state degeneracy, construct the parafermion operators, and study their properties. Finally, in Sec.~\ref{sec:relaxing_J} we relax the strong-backscattering limit in order to study the tunneling between different states in the ground-states manifold and the relation to the $8\pi$ periodicity of the junction.

\section{Review of the Basic Physical Picture and Results}
\label{sec:review}
In this section we aim to cover the basic physics of the proposed setup and present the main results, with as little technical details as possible. These we leave to the more detailed analysis and calculations that will follow in later sections.

\begin{figure}[t]
\includegraphics[width=0.45\textwidth]{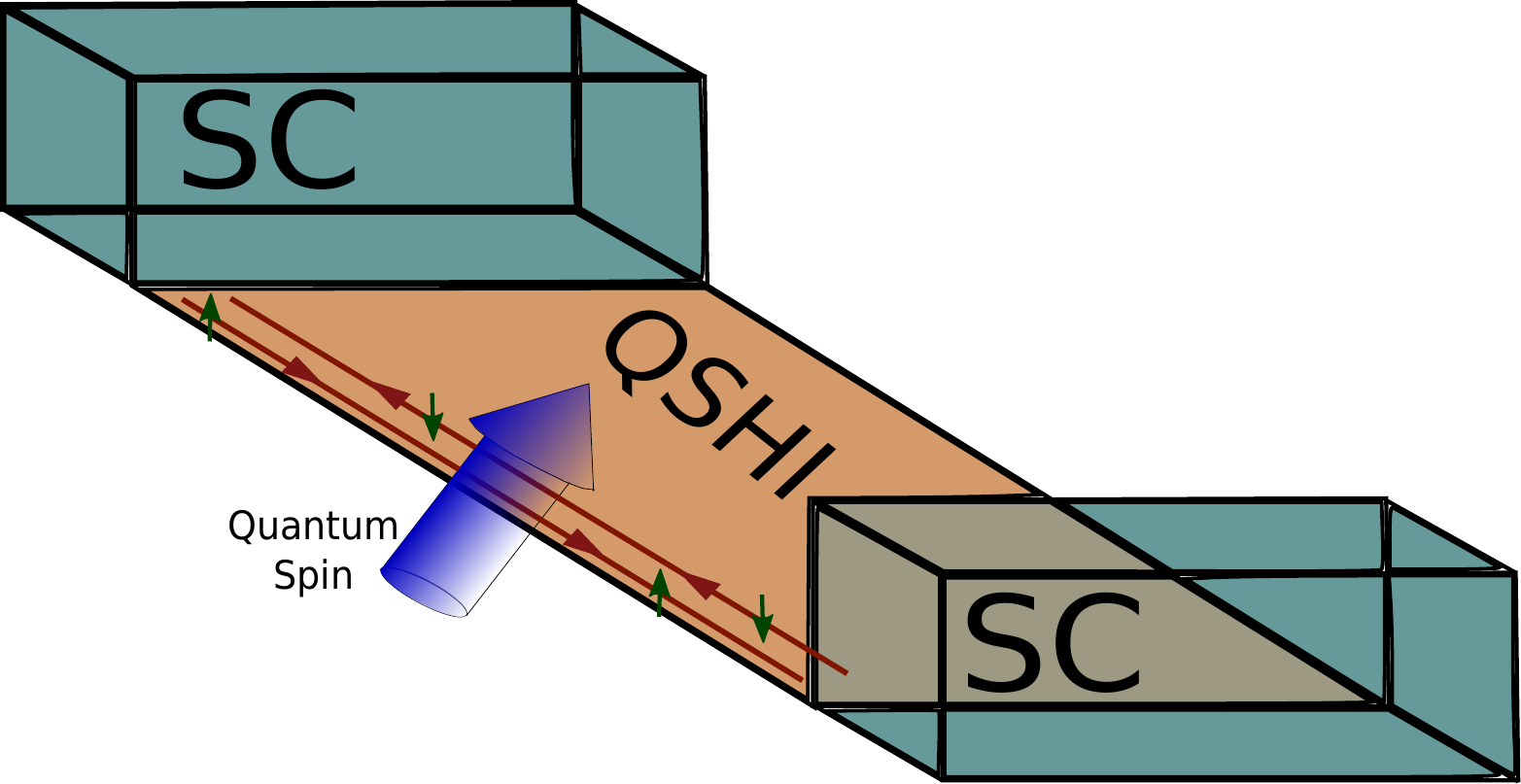}
% \vspace{0pt}
\caption{(color online) Schematic description of the setup. Two superconductors (SC) are connected via a quantum spin Hall insulator (QSHI), with counter propagating chiral electronic modes along its edge. The edge of the QSHI is further coupled to an quantum spin. The channel on the opposite edge was omitted for clarity.}
\label{fig:setup}
\end{figure}

We study a setup consisting of an edge of a quantum spin Hall insulator, with a helical pair of counter-propagating edge modes. The junction connects parts of the edge that are in proximity to BCS superconductors, creating a Josephson junction. In addition to that, the edge modes in the junction are coupled by exchange interaction to an impurity spin. We draw a schematic picture of the setup in Fig.~\ref{fig:setup}. The entire setup is time reversal symmetric, but all other symmetries, in particular spatial symmetries that pertain to the coupling between the edge electrons and the impurity, are assumed to be broken. The lack of spatial symmetry can be the result of the impurity spin representing an accidental charge puddle along the edge~\cite{Vayrynen_2013}, or engineered as part of a setup where the coupling to the impurity is an effective representation of coupling to a quantum dot put in proximity to the edge.

Previous studies of time reversal symmetric topological Josephson junctions without coupling to a spin reported that interactions between the electrons in the junction opens a gap in the many-body spectrum, whose signature is the $8\pi$ periodicity of the current with the phase bias across the junction~\cite{Zhang_2014,Orth_2015}. Strong pair backscattering within the junction favors magnetic order of the edge electrons, and gaps the entire junction. The condition for the magnetic order to appear is that the pair backscattering term is relevant, from a renormalization group point of view. This happens for a sufficiently long junction if the electron-electron interactions are strong such that the Luttinger parameter describing them maintain $g<1/2$.

In junctions with broken time reversal symmetry, a doubly degenerate ground state is related to two possible fermion parities in the junction, and the degeneracy is captured by a pair of Majorana fermions, whose joint occupancy decides the fermion parity~\cite{Fu_2009}. The degeneracy is lifted due to the weak coupling of the Majorana fermions across a finite junction. Maintaining time reversal symmetry augments the degeneracy caused by the fermion parity, as pairs of time reversed states are also degenerate. These states pertain to different magnetic orders in the junction, and their degeneracy gives rise to a four-fold degenerate ground-state space, which is appropriately described by $\mathbb{Z}_4$ parafermions. The degeneracies are lifted as the $\mathbb{Z}_4$ parafermions are weakly coupled, allowing tunneling of fractional $e/2$ charge across the junction, with the weakness of the hybridization ensured by the gap along the entire junction.

Addressing the setup we present here, it is not at all evident that a similar gap can open as a result of coupling to an impurity. While the pair backscattering term acts along the entire length of the junction, the coupling to the impurity is limited to a single point in space. However, previous studies of impurities in Luttinger liquids, most notably those by Kane and Fisher \cite{Kane_1992}, show that the backscattering from an impurity is relevant for any repulsive interaction between the electrons $g<1$, and the backscattering term increases in magnitude under the renormalization group flow. The strong coupling regime is that of infinitely strong backscattering, effectively separating both sides of the impurity into two independent Luttinger liquids.

The exchange interaction between the electrons and the impurity spin is not only subject to the flow described by Kane and Fisher, but also to Kondo physics. In contrast to the strong backscattering regime, in the Kondo strong coupling regime the impurity favors forming a singlet state with the edge electrons, resulting in the complete screening of the impurity and the reconstitution of the edge modes around the impurity site~\cite{Wu_2006,Maciejko_2009}.

The competition between these two strong coupling regimes is decided based on the strength of the electron-electron interactions in the junction, encoded in the Luttinger parameter $g$, and on the bare anisotropy of the exchange couplings. Strong interactions and strong anisotropy will favor the strong backscattering regime, while the Kondo flow tends to erase any initial anisotropy and lead to an isotropic strong coupling fixed point that is characterized by a singlet.

In addition to the exchange couplings, the system is also defined by energy scales pertaining to the junction itself. These are the superconducting gap $\Delta$ and the level spacing $\delta$ of the subgap Andreev states located in the junction. These scales will generally cut the renormalization group flow before it reaches one of the strong coupling fixed points, even at zero temperature. In case that they are the largest energy scales, the exchange coupling will not renormalize and will remain weak, allowing for a perturbative treatment. This regime was studied in detail in Refs.~\cite{Peng_2016} and ~\cite{Hui_2017} and will not be considered here. The cutting of the flow means that the system will reside in the vicinity of one of the strong coupling fixed points (or stay in the weak coupling regime, if $\Delta$ and $\delta$ are large,) and the low-energy physics will be determined by the perturbations within each regime.

\begin{figure}[t]
\includegraphics[width=0.45\textwidth]{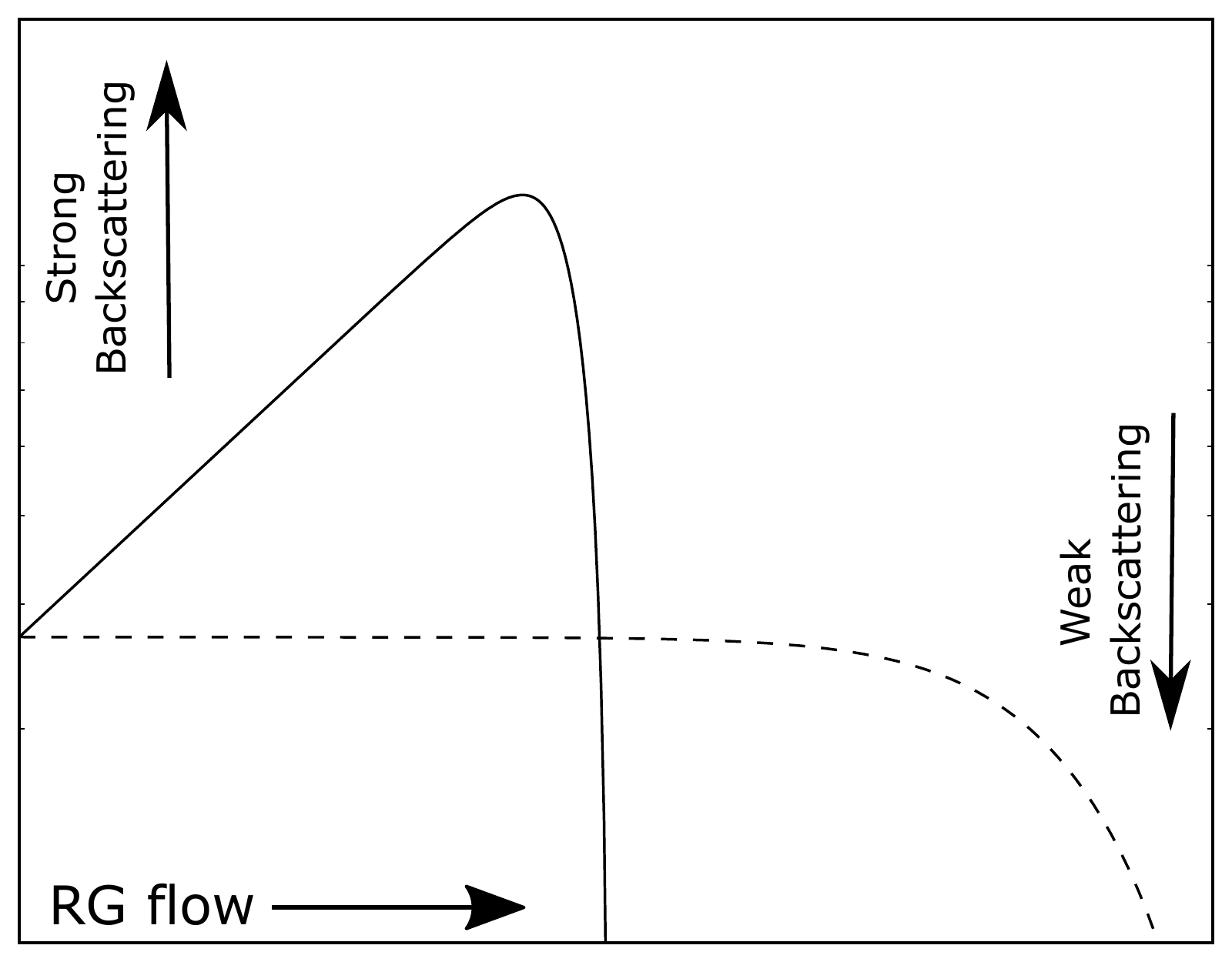}
% \vspace{0pt}
\caption{Schematic description of characteristic renormalization group flow of the backscattering strength for interacting (solid line) and noninteracting (dashed) edge. The strength of the backscattering is plotted on a logarithmic scale.}
\label{fig:schematic_rg}
\end{figure}

\begin{figure}[t]
\includegraphics[width=0.45\textwidth]{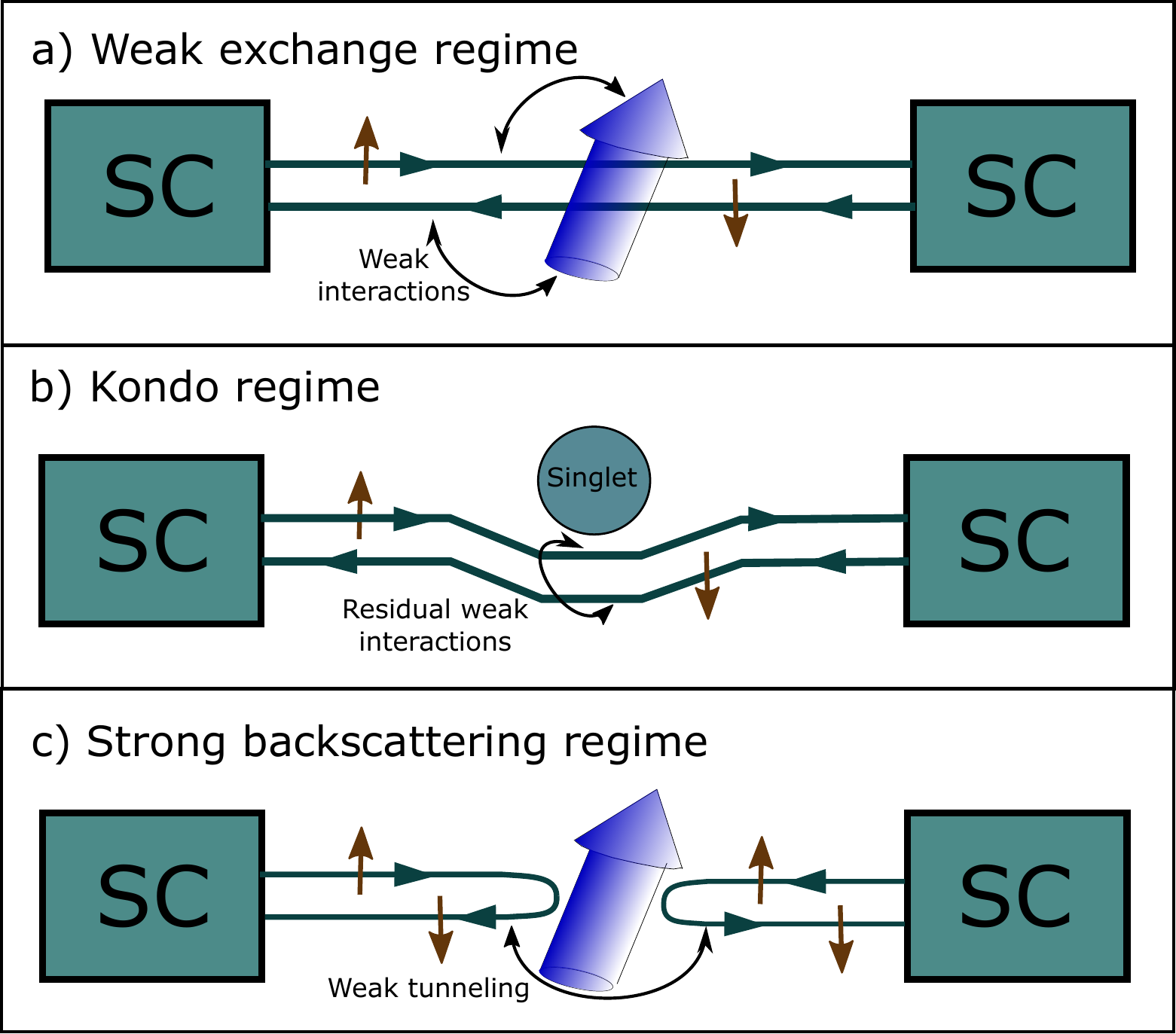}
% \vspace{0pt}
\caption{(color online) The three different regimes discussed in the main text. a) The weak exchange regime, where the exchange couplings are much smaller than the superconducting gap, and the interactions with the impurity spin are treated perturbatively. b) The Kondo regime, where the spin impurity is almost completely screened, and the anisotropic nature of the exchange coupling give rise to residual interactions between the edge electrons. c) The strong backscattering regime, where the junction is effectively cut into two separate parts, weakly interacting with each other.}
\label{fig:diff_limits}
\end{figure}

A schematic picture of a typical renormalization group flow of the backscattering strength $J_{B}$ is given in Fig.~\ref{fig:schematic_rg}. For $g<1$, the relative strength of the backscattering initially increases exponentially at rate $(1-g)$, but as the flow continues the Kondo physics eventually tends to dominate and $J_{B}$ decreases to zero. For the noninteracting case $g=1$ there is no increase and $J_{B}$ will always decrease until the backscattering is suppressed by singlet formation and screening of the impurity. If the renormalization group flow stops at a point where $J_{B}$ is the largest energy scale, the system will reside in the strong backscattering regime. On the other hand, if the flow stops after $J_{B}$ has decreased to small values, the system will be in the Kondo regime. A schematic depiction of all three regimes -- the weak exchange coupling, the Kondo regime, and the strong backscattering regime -- is given in Fig.~\ref{fig:diff_limits}. An exact flow diagram for several different bare values of the exchange coupling and different values of $g$ will be discussed in Sec.~\ref{sec:RG}.

In the strong backscattering regime the setup favors magnetic order of the edge electrons about the impurity. Similar to the strongly interacting regime studied by Zhang and Kane, the boundaries between the magnetic and the superconducting order give rise to $\mathbb{Z}_4$ parafermions, which are located to the left and to the right of the impurity. A unique feature of these parafermions is that they are not composed solely of the edge electrons degrees of freedom but also contain parts that pertain to the impurity spin. These parafermions allow for coherent tunneling of a fractional $e/2$ charge across the junction.

\begin{figure}[t]
\begin{tabular}{l}
(a) \\ \includegraphics[width=0.45\textwidth]{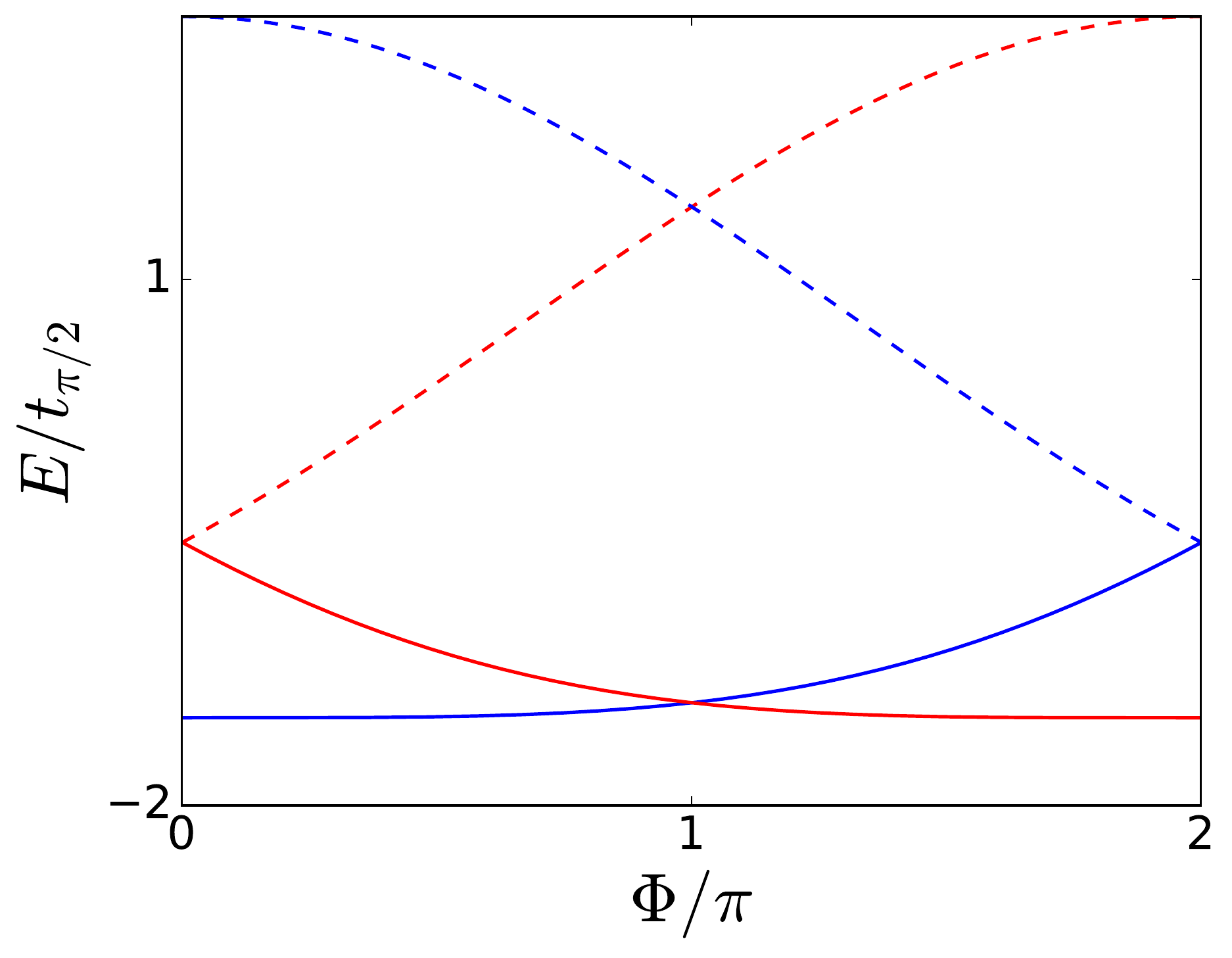}
\end{tabular}
\begin{tabular}{l}
(b) \\ \includegraphics[width=0.45\textwidth]{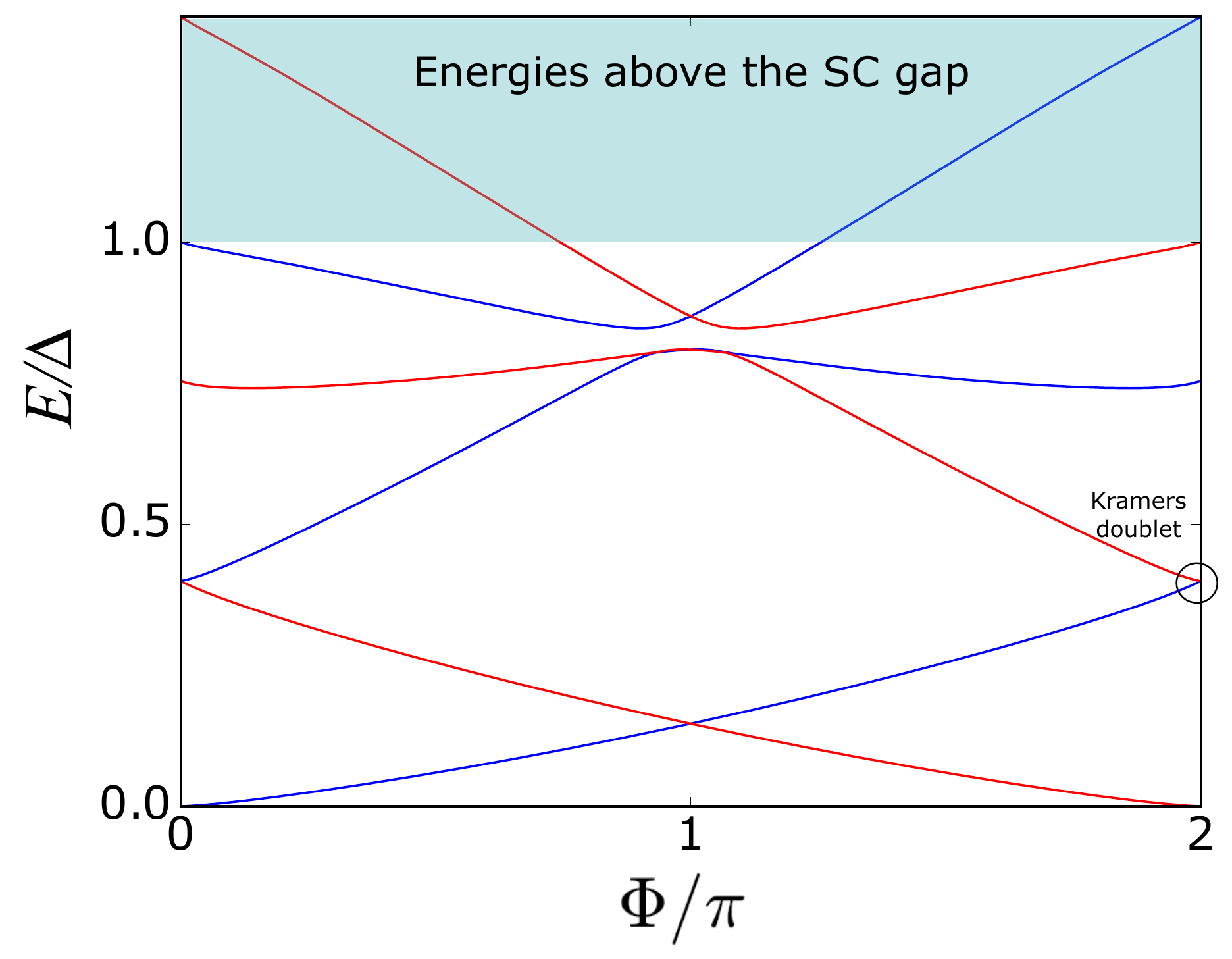}
\end{tabular}
% \vspace{0pt}
\caption{(color online) The many body low energy spectrum as a function of the phase bias $\Phi$ across the junction, for (a) the strong backscattering regime and (b) the Kondo regime with weak residual interactions. In (a), the hopping between levels within the subspace was taken as $t_{pi/2} = 4t_{\pi}$ (see Sec.~\ref{sec:relaxing_J}.) In (b), the energy is given in units of the superconducting gap, and as a gap opens the four lowest lying states are separated from it. The black circle points the protected Kramers doublet due to the time reversal symmetry. Here, the length of the junction was taken as $L/(\pi \Delta v_F) = 1$, in order to ensure that enough Andreev bound states exist and interact in the junction simultaneously. The parameters for the residual exchange coupling and potential scattering are given in App.~\ref{app:Kondo}. For both plots, the $8\pi$ periodicity does not depend on the specific choice of parameters.}
\label{fig:eff_H_energies}
\end{figure}

In the limit of infinite backscattering, the ground-state subspace is four-fold degenerate. Relaxing this limit, the parafermions weakly couple and tunneling between the different states is allowed, lifting the degeneracies. The tunneling terms are exponentially small in $\sqrt{J_{B}/v_F}$. The tunneling is also dependent on the phase bias across the junction, and the energy spectrum as a function of the phase bias is given in Fig.~\ref{fig:eff_H_energies}(a). This dependence gives rise to $8\pi$ periodicity of the current as a function of the phase.

The different states in the ground-state manifold correspond to different fermion parity and different magnetic order. The latter is potentially experimentally accessible by measuring the orientation of the impurity spin, or the appropriate corresponding quantity in a quantum dot that acts as the impurity. A flip in the orientation of the impurity spin is associated with a tunneling of $e/2$ charge across the junction, in a similar manner to the phenomena in which a full rotation of a classical magnet that gaps the edge of a topological spin Hall sample pumps an electron across the gapped region~\cite{Mahfouzi_2010,Meng_2014,Arrachea_2015,Silvestrov_2016}. This phenomenon is present also in the absence of superconductivity, see Refs.\;\cite{Qi_2008,Maciejko_2009}. We believe that this allows the observation and measurement of such parafermions using realistic devices.

As a final comment, we point out that the low-energy physics in the Kondo strong coupling regime is similar to the one in the weak coupling regime studied in Refs.~\cite{Peng_2016} and ~\cite{Hui_2017}. The screened impurity induces weak local spin-spin interactions between the edge electrons themselves. The low-energy many-body spectrum, seen here in Fig.~\ref{fig:eff_H_energies}(b), is once again $8\pi$ periodic as a function of the phase bias. Similarly to the weak coupling regime, the level crossings that give rise to this periodicity are protected by fermion parity conservation and time-reversal symmetry. This strong-to-weak duality is an interesting effect which is covered in Sec.~\ref{sec:RG-Kondo}.

\section{Model and Bosonization}
\label{sec:Model_and_Bos}
We consider a quantum spin Hall edge of length $L$, with edge modes counterpropagating with velocity $v_F$, placed between two superconducting leads with superconducting gap $\Delta$ and phase difference $\Phi$ (see Fig.~\ref{fig:setup}). The helical edge modes are then described by the Hamiltonian $\mathcal{H}_0+\mathcal{H}_{\Delta}$ with~\citen{units}
\begin{eqnarray}
	\mathcal{H}_0 &=& -iv_F\sum_{\sigma=\pm}\sigma \int\! dx 
	\psi^{\dagger}_{\sigma}(x)\partial_x\psi_{\sigma}(x)
	\nonumber \\ && +
	\int\!dx dy \rho(x)V(x-y)\rho(y), \nonumber \\
	\mathcal{H}_{\Delta} &=& \int\! dx \left[\Delta(x) 
	\psi^{\dagger}_{\uparrow}(x)\psi^{\dagger}_{\downarrow}(x)
	+{\rm h.c.}\right],
	\label{eq:H0+HDelta}
\end{eqnarray}
where $\sigma=+1$ $(-1)$ for up- (down-) spins and
\begin{equation}
\Delta(x) = \Delta \Theta(|x|-L/2)
			e^{-i\frac{\Phi}{2}\;{\rm sgn}(x)},
\end{equation}
is the proximity-induced superconducting potential acting only outside the junction region. The effective Coulomb interaction $V(x-y)$ between the electrons on the edge is taken to be short-ranged $V(x-y) = U(x)\delta(x-y)$, and to act between the charge densities
\begin{equation}
	\rho(x) = \sum_{\sigma} :\!\psi_{\sigma}^{\dagger}(x)\psi_{\sigma}	(x)\!:,
\end{equation}
where $\!:\cdots\!:$ denotes normal-ordering of the operators. Generally, the Coulomb interaction is screened underneath the superconducting contacts, so that we assume $U(x)$ to be of the form $U(x) = U\Theta(L/2-|x|)$. For phase bias $\Phi = n\pi$ this edge Hamiltonian is time reversal symmetric, as defined by the transformation 
\begin{equation}
	T\psi_{\sigma}(x)T^{-1} =
	 -\sigma\psi_{-\sigma}(x).
	 \label{eq:T_of_fermions}
\end{equation}

The edge of the quantum spin Hall insulator is generally not clean, with disorder and impurities present. Electronic puddles along the edge may act as an effective impurity spin that will couple to the edge modes~\cite{Vayrynen_2013}. Alternatively, one may also consider a setup where a quantum dot is intentionally put in proximity to the edge in a controllable manner. The coupling to the dot can be described effectively as an exchange coupling with an impurity spin. To account for such a coupling we consider a localized impurity spin $\bf{S}$ at the origin $x=0$, described by the exchange Hamiltonian
\begin{equation}
	\mathcal{H}_S = \sum_{\alpha,\beta}J_{\alpha,\beta}
						s^{\alpha}(0)S^{\beta} + 
		\sum_{\alpha}D_{\alpha}(S^{\alpha})^2.
	\label{eq:H_S_electronic}
\end{equation}
Here $s^{\alpha}(x) = {\bf \Psi}^{\dagger}(x)\sigma^{\alpha}{\bf \Psi}(x)$, with $\sigma^\alpha$ the relevant Pauli matrix and ${\bf \Psi}(x) = \left[\psi_{\uparrow}(x)\;\psi_{\downarrow}(x)\right]^T$ the electronic spinor, is the $\alpha$-component of the electronic spin-density. Moreover, the $D_{\alpha}$ account for spin anisotropy. The coefficients $J_{\alpha,\beta}$ are arbitrary exchange-coupling coefficients which we assume to be random (but fixed.) The arbitrariness of the exchange couplings removes all symmetries besides time reversal symmetry, which is preserved since $T{\bf S}T^{-1}=-{\bf S}$. The main results we shall derive are independent of the magnitude of the spin. However, for definiteness, we shall consider the case of a spin-$1/2$ impurity, and comment where the generalization to higher spins is not immediate. Therefore, we omit the spin anisotropy contributions to $\mathcal{H}_S$ in the meantime.

\subsection{Abelian Bosonization}
Our goal now is to map the Hamiltonian of Eqs.~(\ref{eq:H0+HDelta}) and ~(\ref{eq:H_S_electronic}) onto a bosonized one corresponding to two conjugate degrees of freedom that describe superconducting and magnetic order. To this end, we start by bosonizing the fermionic fields $\psi_{\sigma}(x)$, which according to the standard prescription~\cite{Haldane_1981} can be written as
\begin{equation}
	\psi_{\sigma}(x) = \frac{1}{\sqrt{2\pi a}}e^{-i\phi_{\sigma}(x)},
\end{equation}
where $\phi_{\sigma}(x)$ are bosonic fields and $a$ is a short-distance cutoff associated with the electronic bandwidth $D$ by $a\sim\pi v_F / D$. These bosonic fields obey the commutation relations
\begin{eqnarray}
	\left[\phi_{\sigma}(x),\phi_{\sigma}(y)\right] &=& i\pi\sigma
				{\rm sgn}\{x-y\},
	\nonumber \\ 
	\left[\phi_{\uparrow}(x),\phi_{\downarrow}(y)\right] &=& -i\pi,
\end{eqnarray}
and the identity $:\!\!\psi^{\dagger}_{\sigma}(x)\psi_{\sigma}(x)\!\!: = -\sigma\partial_x\phi_{\sigma}(x)/2\pi$ applies. The transformation of the fields under time reversal is derived from the transformation of the corresponding fermionic fields given in Eq.~(\ref{eq:T_of_fermions}), which dictates
\begin{eqnarray}
	T\phi_{\uparrow}(x)T^{-1} &=& -\phi_{\downarrow}(x)+\pi,
	\nonumber \\
	T\phi_{\downarrow}(x)T^{-1} &=& -\phi_{\uparrow}(x).
\end{eqnarray}

The different terms in the Hamiltonian are now written using these bosonic fields as
\begin{widetext}
\begin{eqnarray}
	\mathcal{H}_0 &=& \frac{v_F}{4\pi}\sum_{\sigma}
	\int\!dx \left[\partial_x\phi_{\sigma}(x)\right]^2 +
	\int\!dx \frac{U(x)}{4\pi^2}\left[\partial_x\phi_{\uparrow}(x)-
	\partial_x\phi_{\downarrow}(x)
	\right]^2, \nonumber \\
	\mathcal{H}_\Delta &=& 
	\int\!dx\frac{\Delta}{2\pi a}\Theta\left(|x|-\frac{L}{2}\right)
	\sin\left[\phi_{\uparrow}(x)+\phi_{\downarrow}(x)-
	\frac{\Phi}{2}{\rm sgn}\{x\}\right],
\end{eqnarray}
and we divide $\mathcal{H}_S$ into backscattering and forward scattering terms $\mathcal{H}_S = \mathcal{H}_{B} + \mathcal{H}_{F}$ with~\citen{couplings}
\begin{eqnarray}
	\mathcal{H}_{B} &=& \sum_{\beta} 
	\left\{\frac{J_{x,\beta}}{\pi a}
	\sin\left[\phi_{\uparrow}(0)-\phi_{\downarrow}(0)\right]
	-\frac{J_{y,\beta}}{\pi a}
	\cos\left[\phi_{\uparrow}(0)-\phi_{\downarrow}(0)\right]
	\right\} S^{\beta}, \nonumber \\
	\mathcal{H}_{F} &=& -\sum_{\beta}\frac{J_{z,\beta}}{2\pi}
	\left[\partial_x\phi_{\uparrow}(0)+\partial_x\phi_{\downarrow}(0) 
	\right]S^{\beta}.
	\label{eq:H_bos_s}
\end{eqnarray}
\end{widetext}

It is natural to introduce the linear combinations $\varphi(x) = [\phi_{\uparrow}(x)+\phi_{\downarrow}(x)]/2$ and $\theta(x) = [\phi_{\uparrow}(x)-\phi_{\downarrow}(x)]/2$. These fields are a conjugate pair, as they obey the commutation relations $$[\varphi(x),\varphi(y)]=0=[\theta(x),\theta(y)]$$ and
\begin{equation}
	\left[\varphi(x),\theta(y)\right] = i\pi\Theta(x-y).
	\label{eq:phi_theta_comm}
\end{equation}
The derivatives of these fields give the physical charge-density $\rho(x)=e\partial_x\theta(x)/\pi$ and spin-$z$-density $\sigma^z(x)=-\partial_x\varphi(x)/\pi$, with $(-e)$ the charge of the electron. Under time reversal these fields transform as
\begin{eqnarray}
	T\varphi(x)T^{-1} &=& -\varphi(x)+\pi/2, \nonumber \\
	T\theta(x)T^{-1} &=& \theta(x)+\pi/2.
\end{eqnarray}

These fields allow us to write the Hamiltonian in a more compact form
\begin{eqnarray}
	\mathcal{H}_0 &=& \frac{v_F}{2\pi}
	\int\! dx
	\left\{\left[1+\frac{U(x)}{2 \pi v_F}
	\right]\left[\partial_x\theta(x)\right]^2 + 
	\left[\partial_x\varphi(x)\right]^2\right\},
	\nonumber \\
	\mathcal{H}_{\Delta} &=&
	\int\!dx\frac{\Delta}{\pi a}
	\Theta\left(|x|-\frac{L}{2}\right)
	\sin\left[2\varphi(x)-\frac{\Phi}{2}{\rm sgn}\{x\}\right],
	\nonumber \\
	\mathcal{H}_{B} &=& \left\{
	\frac{{\bf  J}_{x}}{\pi a}\sin[2\theta(0)]-
	\frac{{\bf  J}_{y}}{\pi a}\cos[2\theta(0)]
	\right\}\cdot {\bf  S},
	\nonumber \\
	\mathcal{H}_{F} &=& -\frac{{\bf  J}_{z}}{\pi}\partial_x\varphi(0)
		\cdot {\bf  S},
	\label{eq:H_bos_before_RG}
\end{eqnarray}
where we have adopted a vector notation for the coupling constants ${\bf  J}_\alpha = \sum_{j} J_{\alpha,\beta} \hat{x}_\beta$. Assuming $U(x)$ to be constant in space throughout the junction, $U(x)=U$, it defines the Luttinger parameter $g = (1+U/2\pi v_F)^{-1/2}$, with $g<1$ corresponding to repulsive interactions and $g>1$ to attractive interactions. The free part of the Hamiltonian inside the junction is then written in the usual manner as $$\mathcal{H}_0 = \frac{v_F}{2\pi g}\int\!dx [\frac{1}{g}(\partial_x\theta)^2+g(\partial_x\varphi)^2].$$ It is the Hamiltonian of Eq.~(\ref{eq:H_bos_before_RG}) that will be our main point of interest in most of the paper.

\section{Perturbative Renormalization Group Analysis}
\label{sec:RG}
In order to study the low-energy properties of the model, we apply a perturbative renormalization group analysis to it. For clarity and brevity, we relegate the detailed calculations to App.~\ref{app:RG} and present here their results.

\subsection{Weak Coupling Fixed Point}
The leading order renormalization group flow equations about the weak coupling fixed point follow from the results of Kane and Fisher\cite{Kane_1992} on impurities in Luttinger liquids. These results describe relevant (irrelevant) backscattering for repulsive (attractive) interactions. The leading order equations are given by
\begin{eqnarray}
	\frac{d{\bf  J}_{\alpha}}{dl} &=& (1-g){\bf  J}_{\alpha}
	\;\;\;\; {\rm for} \; \alpha=x,y, \nonumber \\
	\frac{d{\bf  J}_{z}}{dl} &=& 0,
	\label{eq:RG_bosonic_leading_order}
\end{eqnarray}
where $l=-\ln(D'/D)$, with $D$ the bare bandwidth and $D'$ the running bandwidth scale, and we assumed a uniform Luttinger parameter $g$ throughout the junction. In addition, the superconducting gap $\Delta$, assuming that the interactions are  screened in the superconducting region, obeys the flow equation $d\Delta/dl = \Delta$.

These results are in agreement with previously known results for magnetic impurities in Luttinger liquids~\cite{Furusaki_1994,Schiller_1995,Andergassen_2006,
Furusaki_2005,Vayrynen_2016}. For any repulsive interaction $g<1$ the backscattering Hamiltonian $\mathcal{H}_{B}$ is relevant, flowing to the strong coupling regime, and irrelevant for attractive interactions. The forward scattering $\mathcal{H}_{F}$ is marginal to leading order. Finally, the induced superconducting gap $\Delta$ is relevant.

The exchange coupling also leads to Kondo behavior, captured by the second-order terms in the flow equations. These terms are of the form (see App.~\ref{app:RG})
\begin{eqnarray}
	\frac{d{\bf  J}_i}{dl}= \cdots + \frac{C_i}{2v_F} \epsilon_{ijk} 
				{\bf  J}_j \times {\bf  J}_k,
	\label{eq:RG_bosonic_second_order},
\end{eqnarray}
where the $\cdots$ on the right-hand-side stand for the linear terms in Eqs.~(\ref{eq:RG_bosonic_leading_order}). The second-order perturbation also gives rise to new terms that represent two-body interactions at the origin, such as $\cos[4\theta(0)]$, $[\partial_x\varphi(0)]^2$ {\it etc.} These terms are irrelevant near $g=1$, and we omit them. We note that for $g<1/4$ the pair backscattering may play an important role, as it becomes relevant~\cite{Fabrizio_1995,Egger_1998,Furusaki_2005,Maciejko_2009}. For interactions of this strength, one also has to consider the bulk pair-backscattering which is relevant for $g<1/2$. Such strong values of $g$ were discussed in Refs.~\cite{Zhang_2014} and ~\cite{Orth_2015} and we will not consider them here. The coefficients $C_i$ in Eq.~(\ref{eq:RG_bosonic_second_order}) are of order unity and depend on the cutoff scheme. For a hard cutoff of momenta at $\Lambda$, associated with the energy bandwidth $D$, they are given by
\begin{eqnarray}
	C_z &=& \frac{4\pi g^3}{(\pi a \Lambda)^2}
	{\rm Im}\left\{
	\int_0^{\pi/2}\! dx  x e^{ix-2gJ(x)}
	\right\}
	, \nonumber \\
	C_x &=& C_y = 2g,
\end{eqnarray}
where
\begin{equation}
	J(x) = \int_{0}^{1}\!\frac{ds}{s}\left[1-e^{ixs}\right],
	\label{eq:J_defined}
\end{equation}
which results in positive $C_z$ near $g=1$. To corroborate this picture we also conducted a poor man's scaling analysis to the original electronic Hamiltonian in the noninteracting regime, reaching equivalent equations to second order in the exchange coupling.

We stress here that while the results of the leading order are independent of the magnitude of the spin-impurity $S$, in deriving the second order terms we explicitly restricted ourselves to $S=1/2$. Larger impurity spins will lead to modified flow equations in the second-order, which will include corrections to the spin anisotropy terms $D_{\alpha}$ as well.

The linear terms in the flow equations increase the exchange couplings ${\bf J}_x$ and ${\bf J}_y$ at the same exponential rate in $l$, driving both to a strong backscattering fixed point while maintaining their orientation and the ratio between them. The second order terms, however, drive the flow toward the Kondo fixed point, characterized by an isotropic exchange coupling ${\bf J}_i \perp {\bf J}_j$ and $|{\bf J}_i|=|{\bf J}_j|$ for all $i\neq j$. At this point the exchange interaction is proportional to $J{\bf s}(0)\cdot{\bf S}$, and the ground state is a singlet.

To see this, we set $g=1$, thus taking into account only the second order terms. We identify the three constants of motion $C_{ij} \equiv {\bf J}_i \cdot {\bf J}_j = |{\bf J}_i||{\bf J}_j|\cos(\theta_{ij})$. As the magnitudes of all ${\bf J}_i$ increase, the cosine must decrease to zero. Therefore $\theta_{ij} \to \pi/2$, and we conclude that all the exchange coupling vectors become orthogonal to each other. Similarly ${\bf J}_x^2-{\bf J}_y^2$ is also a constant of motion, hence the difference in magnitude between the different exchange couplings becomes negligible in the strong coupling regime.

Strong backscattering is the result of pinning $\theta(0)$ to a value which minimizes the energy of $\mathcal{H}_B$ in Eq.~(\ref{eq:H_bos_before_RG}). At the isotropic point, however, $\theta(0)$ serves as a rotation angle and can be eliminated from $\mathcal{H}_B$ by a unitary transformation, corresponding to the absence of backscattering. To quantify the strength of the backscattering, we examine the dependence of the eigenenergies $E_{\pm}$ of $\mathcal{H}_{B}$ in Eq.~(\ref{eq:H_bos_before_RG}) on $\theta(0)$. For a spin-$1/2$ impurity, the energies are given by $E_{\pm} = \pm B[\theta(0)]$ with
\begin{eqnarray}
	B^2[\theta(0)] &=& \frac{{\bf J}_x^2+{\bf J}_y^2}{2(\pi a)^2}-
		\frac{{\bf J}_x^2-{\bf J}_y^2}{2(\pi a)^2}\cos[4\theta(0)]-
		\nonumber \\ &&
		\frac{{\bf J}_x \cdot {\bf J}_y}{(\pi a)^2} \sin[4\theta(0)].
		\label{eq:B_of_theta}
\end{eqnarray}
At the isotropic point $B[\theta(0)]$ becomes completely independent of $\theta(0)$. Following this, we can characterize the magnitude of the anisotropy by
\begin{equation}
	J_{B} =
	\sqrt{\frac{\left({\bf J}_x^2-{\bf J}_y^2\right)^2
	+4\left({\bf J}_x\cdot{\bf J}_y\right)^2}
	{{\bf J}_x^2+{\bf J}_y^2}},
	\label{eq:J_BS}
\end{equation}
which is zero at the isotropic point and increases with the anisotopry.

Under the linear terms of the flow equations, $J_{B}$ will grow exponentially for repulsive interactions, as $J_{B}(l) = J_{B}(0)\exp[(1-g)l]$. Under the second order terms, however, the numerator in the expression for $J^2_{B}$ is a constant of motion, while the denominator grows, and $J_{B}$ will decrease. As two extreme cases one may consider on the one hand a setup in which only ${\bf J}_x \neq 0$, and the other exchange couplings vanish. In this case, the Kondo flow is absent and the backscattering increases exponentially. On the other hand, starting from the perfectly isotropic point where $J_{B}=0$, the system will remain isotropic throughout the renormalization process and no dependence on $\theta(0)$ will emerge.

\begin{figure}[t]
\includegraphics[width=0.45\textwidth]{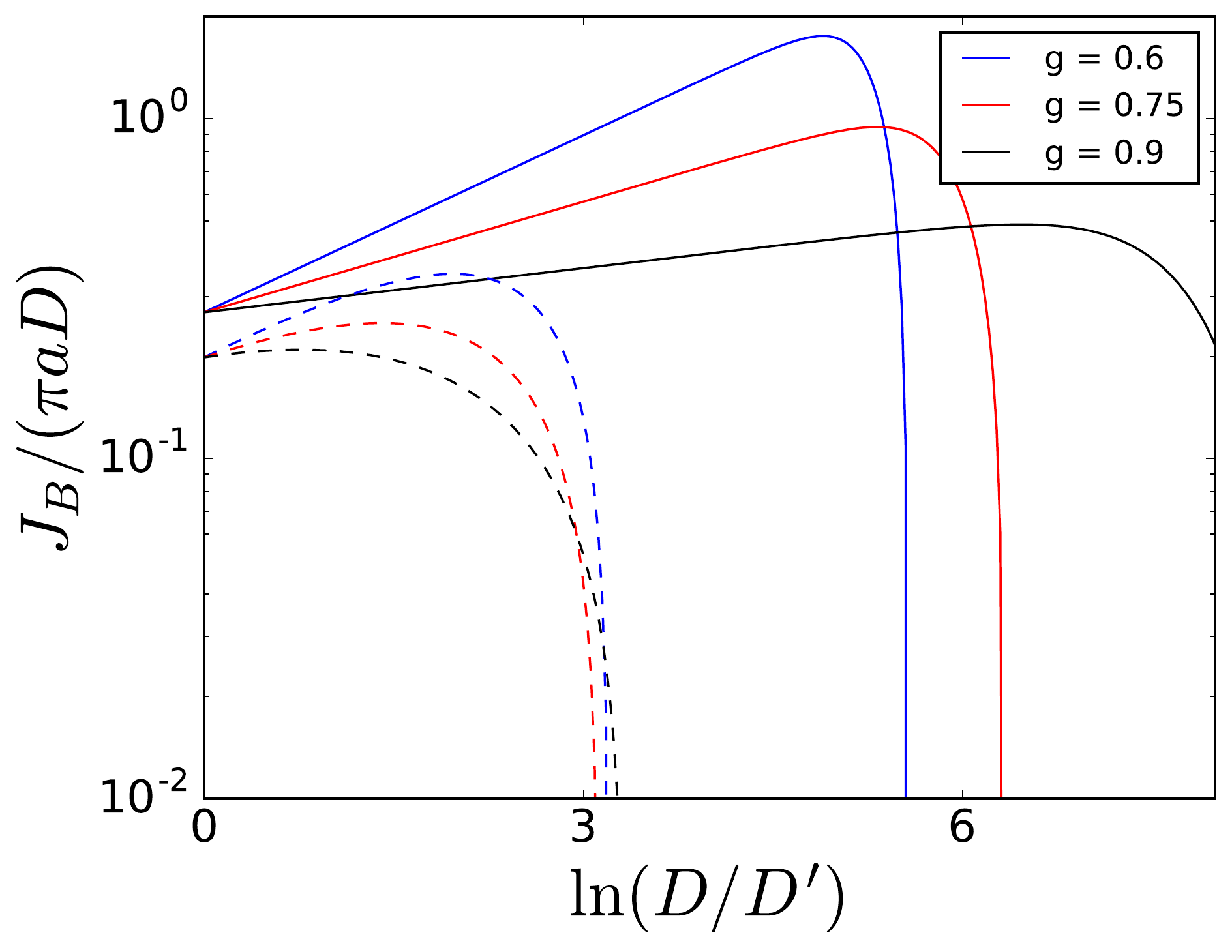}
% \vspace{0pt}
\caption{(color online) The renormalization group flow of the energy scale $J_{B}/(\pi a)$ given in Eq.~(\ref{eq:J_BS}) for two different sets of bare exchange couplings and for different strengths of electron-electron interactions in the junction. The continuous lines correspond to bare value of $J_{B}/(\pi a D) \simeq 0.27$ while the dashed lines correspond to $J_{B}/(\pi a D) \simeq 0.16$. The values of all $9$ exchange couplings for each case are given in App.~\ref{app:RG}. It should be stressed here that the flow cannot be derived only from the bare value of $J_{B}$ and $g$, but a full knowledge of all of the exchange couplings is needed.}
\label{fig:RG_flow_examp}
\end{figure}

The competition between these two directions determines whether the strong-coupling regime that the system will find itself in at the end of the renormalization group flow will be characterized by strong anisotropic backscattering or by Kondo screening of the impurity. This competition is decided by the values of the initial bare anisotropy, the strength of the interactions encoded in $g$ and finally, by the point at which the renormalization flow stops, which depends also on the junction's energy scales $\delta$ and $\Delta$ and on the bandwidth $D$. In Fig.~\ref{fig:RG_flow_examp} we plot several examples for the flow of $J_{B}$ as a function of $l=\ln(D/D')$ under different initial conditions and values of $g$. Initially increasing exponentially, $J_{B}$ generally starts decreasing at some point once the second order terms become dominant. However, it may attain a relatively large value before it starts to decrease, and the flow may stop at that point. As the perturbative flow equations are strictly applicable only in the weak coupling regime, we note that this is a qualitative picture, and a more precise analysis calls for numerical tools, which are beyond the scope of this work. However, as we shall show below, both regimes are stable. 

Below we discuss these two possible strong coupling regimes, and describe their low-energy physics. We will address briefly the Kondo strong coupling regime, before devoting the main part of the discussion to the strong backscattering regime, which gives rise to $\mathbb{Z}_4$ parafermions.

\subsection{Kondo screening regime}
\label{sec:RG-Kondo}
In the strong Kondo coupling regime, the renormalization group flow of the exchange couplings tends to form an isolated singlet of the impurity spin and the adjacent electrons of the edge. The helical nature of the edge does not allow for backscattering from the pure singlet, and the edge will reconstitute itself around the singlet~\cite{Wu_2006,Maciejko_2009}. However, the presence of the junction's energy scales $\delta$ and $\Delta$ mean that the system will not be exactly at the Kondo fixed point even at zero temperature, as the flow will be cut before it reaches that point, and the irrelevant operators about the fixed point will determine the low-energy behavior of the system.

Even though the exchange couplings at the fixed point are isotropic, the anisotropic nature of the bare exchange couplings will have a weak residual effect on the edge electrons. This can be seen by carrying out a strong-coupling expansion about the singlet state with the anisotropic terms as perturbations. The leading terms in that expansion, after projecting onto the singlet, can be written as
\begin{eqnarray}
	V' &=& \lambda_{PS}
		 {\bf \Psi^{\dagger}}(0){\bf \Psi}(0)
		\nonumber \\ && +
		\sum
		\lambda_{\alpha,\beta}
		{\bf \Psi^{\dagger}}(0)\sigma^{\alpha}
		\partial_x{\bf \Psi}(0)
		\partial_x{\bf \Psi^{\dagger}}(0)
		\sigma^{\beta}{\bf \Psi}(0)
		\nonumber \\ && + \cdots,
		\label{eq:Kondo_potential}
\end{eqnarray}
where time reversal symmetry constrains the coefficients $\lambda_{\alpha,\beta}$ to be real and to satisfy $\lambda_{\alpha,\beta}=\lambda_{\beta,\alpha}$, and we have used point-splitting in order to avoid violation of the Pauli principle~\cite{Maciejko_2009}.

In terms of the bosonic fields the first term will be proportional to $\partial_x\theta(0)$ and is marginal for all $g$ to leading order. The second term will consist of operators involving four bosonic fields, and their derivatives, such as  $\cos^2[2\theta(0)][\partial_x\theta(0)]^2$ {\it etc.} These operators are irrelevant for any $g$, and can be treated as weak perturbations about the fixed point.

As in the weak coupling limit~\cite{Peng_2016,Hui_2017}, it is constructive to work in the electronic picture, effectively taking the noninteracting limit, and projecting onto the low-energy space spanned by the Andreev bound states. The finite dimensional Hamiltonian is then diagonalized exactly for any phase-bias $\Phi$, and the spectrum is plotted. We carry out the detailed calculation in App.~\ref{app:Kondo}, and present here only the result, shown in Fig.~\ref{fig:eff_H_energies}(b). For a long enough junction the exchange interaction opens gaps in the many-body spectrum, leading to $8\pi$ periodicity of the Josephson current. The strong-to-weak duality can be understood in the following manner. The screening of the impurity decouples it from the edge, but the screening is not perfect and leaves residual weak spin-spin interactions. These weak interactions have a similar effect to the ones considered in Ref.~\cite{Zhang_2014}.

\subsection{Strong backscattering regime}
\label{sec:rg_strong_b}
For $g<1$, the backscattering terms in the Hamiltonian of Eq.~(\ref{eq:H_bos_before_RG}) are relevant to leading order, while the forward scattering term is marginal. Depending on the bare values of the exchange couplings, the renormalization group flow might stop at a point where the largest energy scale is the backscattering term. In order for this to be the case, the setup must have a strong bare anisotropy in the exchange couplings.

To simplify the discussion in this section, we shall consider the strong local backscattering Hamiltonian to be of the form $\mathcal{H}_{B} = -B\sin[2\theta(0)]S^z$, corresponding to the electronic term $\sigma^x(0)S^z$. This, in fact, is not a restrictive choice, as the basis for the impurity spin can always be chosen such that this local term will be parallel to $S^z$, and we can also apply a rotation about the $z$-axis of the edge electrons' spin, such that the backscattering term will be in the $x$-direction. This is tantamount to a constant shift of the $\theta$ field, which does not affect the kinetic term.

In the limit of infinitely strong backscattering $B\to\infty$, the junction is cut into two disconnected halves by the impurity~\cite{Kane_1992}. The $\theta$ field at the origin is pinned by the backscattering term, and its value is related to the orientation of the impurity spin. The presence of the superconductors makes $\theta(x)$ defined modulo $2\pi$, which renders the ground state four-fold degenerate, as $|\theta(0) = \pi/4, S^z = S\rangle$, $|3\pi/4, -S\rangle$, $|5\pi/4, S\rangle$ and $|7\pi/4, -S\rangle$ all have the same energy $(-B)$, ($S$ is the magnitude of the impurity spin.) This four-fold degeneracy of the ground state is not accidental, but directly related to the time reversal symmetry of the system. Indeed, under a time reversal transformation, $\theta(x) \to \theta(x)+\pi/2$ and ${\bf  S} \to -{\bf  S}$, which relates the different ground states to each other. As the field at $\theta(0)$ is pinned to one of four possible values, we may write it as
\begin{equation}
	\theta(0) = \frac{\pi}{4} + \frac{\pi}{2}\hat{m},
	\label{eq:theta_0_m}
\end{equation}
with $\hat{m}$ an integer-valued operator. The parts of the edge to the left and to the right of the impurity are now described by independent fields $\varphi_{L,R}(x)$ and $\theta_{L,R}(x)$, with the constraint $\lim_{x\to 0}\theta_{L,R}(x) = \theta(0)$.

To establish the stability of the strong backscattering fixed point, we carry out a strong-coupling expansion about it, by considering the local exchange and the kinetic terms as perturbations. Projecting onto the ground-state manifold we get that the leading terms are given by
\begin{eqnarray}
	V' &=& \lambda_1 \cos(\Delta\varphi)+
		\lambda_2\sin(\Delta\varphi)S^z
		\nonumber \\ && +
		\lambda_3
		\partial_x[\theta_R(0^+)-\theta_L(0^-)],
	\label{eq:eff_V_strong_back}
\end{eqnarray}
where $\Delta\varphi = \varphi_R(0^+)-\varphi_L(0^-)$ is the dynamical term connecting both sides of the impurity, and the different coupling coefficients are determined by the values of $J_{\alpha,\beta}$ and $B$.

The terms in Eq.~(\ref{eq:eff_V_strong_back}) can be understood by considering the different constraints on the expansion. As $\theta(0)$ is pinned, the local dynamical degrees of freedom are $\varphi_{L,R}(0^{\pm})$ and the spin. As such, the only time-reversal invariant local perturbations are (i) $\cos(\Delta\varphi)$, which describes tunneling of an electron between the two liquids, (ii) $\sin(\Delta\varphi){\bf S}$ and $\cos(\Delta\varphi)\sin[2\theta(0)]{\bf S}$, which describe tunneling via interaction with the impurity spin, and (iii) the potential scattering $\partial_x\theta_{L,R}(0^{\pm})$. The projection onto the ground state allows for spin flips of the impurity only when they are combined with shifting of $\theta(0)$ by half-integer multiples of $\pi$. As all local processes describe tunneling of full electrons, they change the value of $\theta(0)$ by integer multiples of $\pi$, and therefore cannot connect states with different values of the local spin. The projection onto the ground-state manifold leaves only the $z$-component of $\sin(\Delta\varphi) {\bf S}$, and casts $\cos(\Delta\varphi)\sin[2\theta(0)]{\bf S}$ as identical to $\cos(\Delta\varphi)$. We shall see below that the terms that connect states with different spin orientations are nonperturbative.

The renormalization group flow equations for the operators in Eq.~(\ref{eq:eff_V_strong_back}) are
\begin{eqnarray}
	\frac{d\lambda_j}{dl} &=& \left(1-\frac{1}{g}\right)\lambda_j 
	\;\;\; {\rm for} \; j = 1,2 \nonumber \\
	\frac{d\lambda_3}{dl} &=& 0
\end{eqnarray}
where $l=-\ln(D'/D)$ is the running cutoff. For repulsive interactions $g<1$ the operators $\lambda_{1,2}$ are irrelevant and $\lambda_3$ is marginal, therefore the fixed point is stable under local perturbations. We now turn to study the low-energy physics in the vicinity of the strong backscattering fixed point, with an emphasis on the interplay with the superconductivity.

\section{Infinite backscattering limit and $\mathbb{Z}_4$ Parafermions}
\label{sec:Strong_J}
\begin{figure}[t]
(a)\includegraphics[width=0.45\textwidth]{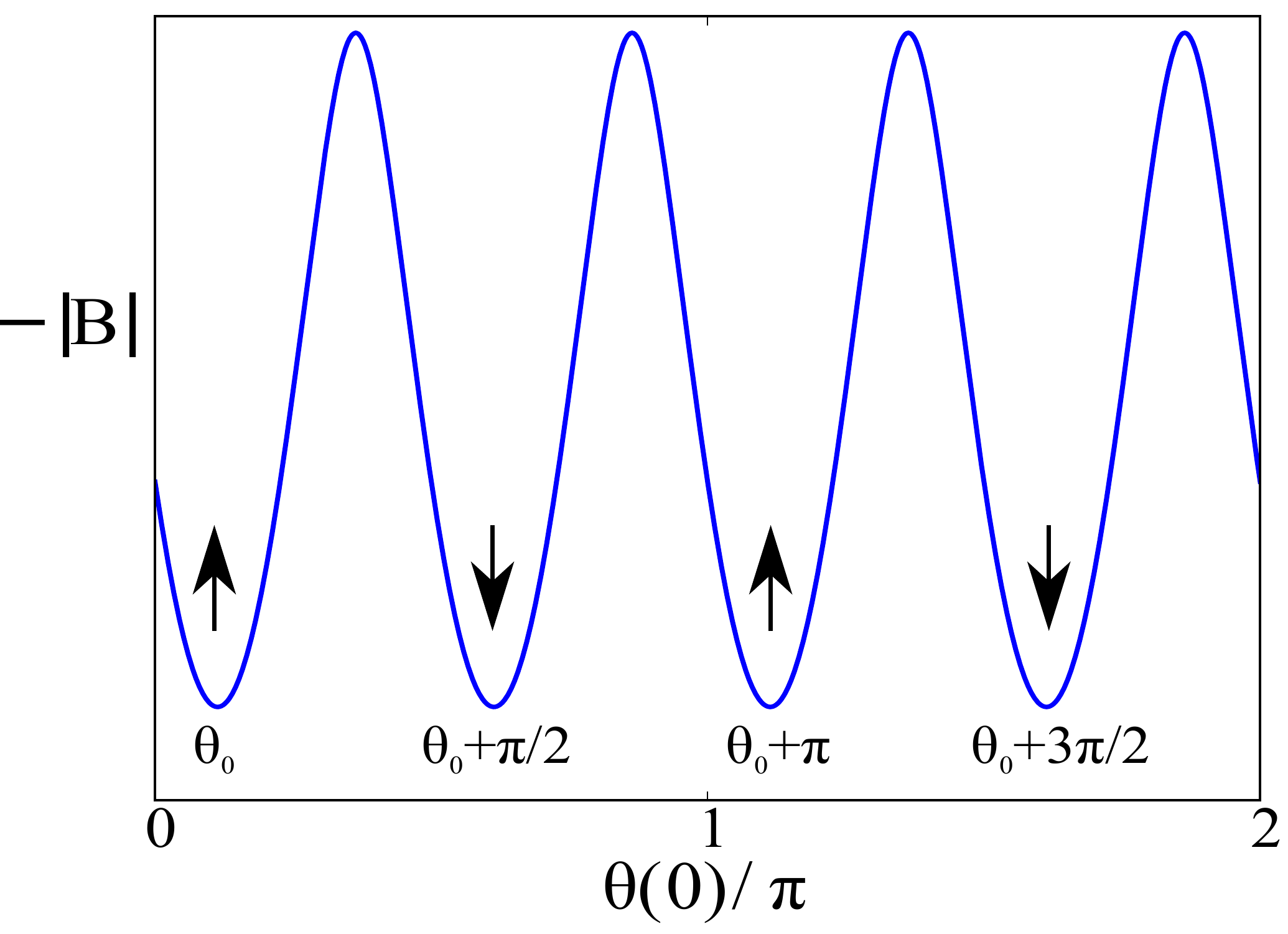}
(b)\;\includegraphics[width=0.44\textwidth]{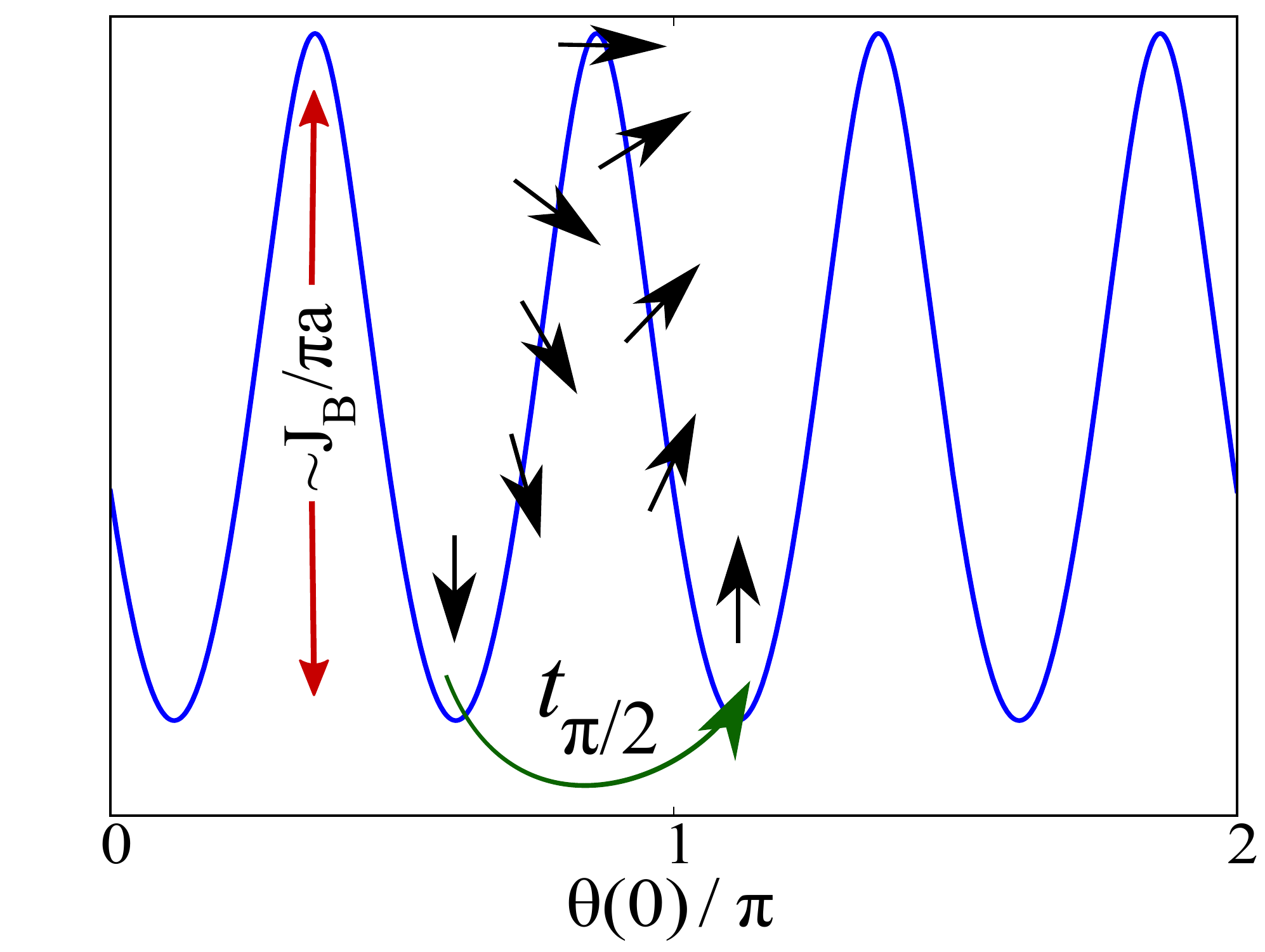}
% \vspace{0pt}
\caption{(color online) (a) Schematic depiction of the magnitude of the lowest energy of $\mathcal{H}_B$ as a function of the field at $\theta(0)$. The values for which the local Hamiltonian is at the ground state are denoted by $\theta_0+m\pi/2$ and $\mathcal{H}_{B}$ pins the field to these minima. The orientation of the impurity spin (black arrows) alternates between adjacent ground states. (b) The instanton process of the tunneling of $\theta(0)$ through a potential barrier of height $\sim J_B/(\pi a)$, that weakly couples two adjacent states within the ground state manifold. The tunneling process is accompanied by the flipping of the impurity spin, which here is schematically depicted as following the effective magnetic field of Eq.~(\ref{eq:eff_magnetic_field}).}
\label{fig:low_e}
\end{figure}

We shall focus now on the regime where the exchange coupling to the impurity spin causes infinitely strong backscattering, effectively cutting the junction into two halves. As discussed before, $\theta(0)$ is fixed by the strong backscattering, and further generalizing Eq.~(\ref{eq:theta_0_m}) we write
\begin{equation}
	\theta(0) = \pi \frac{\hat{m}}{2} + \theta_0,
	\label{eq:theta_0_def}
\end{equation}
where $\theta_0$ is fixed and defined by the exact form of the backscattering terms such that $B[\theta(0)]$, given in Eq.~(\ref{eq:B_of_theta}), is maximal at $\theta(0)=\theta_0$. The dependence of the potential on $\theta(0)$ and the impurity spin orientation can be seen in  Fig.~\ref{fig:low_e}(a).

Outside the junction, the superconducting pairing pins $\varphi(x)$ to the minimum of the potential $\sin[2\varphi(x) - (\Phi/2) {\rm sgn}(x)]$ and we may expand it similarly to $\theta(0)$ as
\begin{equation}
	\varphi\left(\pm\frac{L}{2}\right) = \pi\hat{n}_\pm \pm \frac{\Phi}{4},
	\label{eq:n_pm}
\end{equation}
where $\hat{n}_\pm$ are integer valued operators as well. It should be pointed out that $\hat{m}$ and $\hat{n}_+$ do not commute, $\left[\hat{m},\hat{n}_+\right]=2i/\pi$, and the system cannot be in an eigenstate common to both operators. On the other hand, $\hat{m}$ and $\hat{n}_-$ commute, see Eq.~(\ref{eq:phi_theta_comm}).

The four-fold degeneracy of the ground state gives rise to parafermion operators, associated with the tunneling between the different ground states. The parts of the junction to the left and to the right of the impurity are then domain walls between a region with superconducting order ($\varphi$ is pinned) and magnetic order ($\theta$ is pinned). To derive the corresponding operators we follow a procedure similar to Ref.~\cite{Clarke_2013}. The free Hamiltonian on the left and on the right hand side of the impurity is written as $\mathcal{H}_0=\mathcal{H}_{0,L}+\mathcal{H}_{0,R}$ with
\begin{equation}
	\mathcal{H}_{0,L/R} = \frac{v_F}{2\pi g}
	\int_{-\frac{L}{2}/0}^{0/\frac{L}{2}}\! dx 
	\left[\frac{1}{g}(\partial_x\theta_{L,R})^2
	+g(\partial_x\varphi_{L,R})^2\right],
\end{equation}
and the fields are subject to the boundary conditions
\begin{eqnarray}
	\varphi_L\left(-\frac{L}{2}\right) &=& 	
	\pi \hat{n}_- - \frac{\Phi}{4},
	\nonumber \\
	\varphi_R\left(\frac{L}{2}\right) &=& 									\pi \hat{n}_+ + \frac{\Phi}{4},
	\nonumber \\
	\theta_{L,R}(0) &=& \frac{\pi}{2} \hat{m} + \theta_0.
\end{eqnarray}
These fields can be expanded in their eigenmodes separately in the sections to the left of the impurity as
\begin{eqnarray}
	\varphi_{L}(x) &=& \sum_{k\geq 0}\sqrt{\frac{2\pi}{g\lambda_k L}} 
				\sin\left[\lambda_k\left(x+\frac{L}{2}\right)\right]
	i\left(a_{L,k}-a^{\dagger}_{L,k}\right)
	\nonumber \\ && +\varphi\left(-\frac{L}{2}\right)
	, \nonumber \\
	\theta_{L}(x) &=& \sum_{k\geq 0} \sqrt{\frac{2\pi g}{\lambda_k L}}
				\cos\left[\lambda_k\left(x+\frac{L}{2}\right)\right]
	\left(a_{L,k}+a^{\dagger}_{L,k}\right)
	\nonumber \\ && +\theta(0),
	\label{eq:mode_exp_left}
\end{eqnarray}
and to the right of the impurity as
\begin{eqnarray}
	\varphi_{R}(x) &=& \sum_{k\geq 0}
	\sqrt{\frac{2\pi}{g\lambda_k L}}
				\cos\left(\lambda_k x\right)
	\left(a_{R,k}+a_{R,k}^{\dagger}\right)
	\nonumber \\ && +\varphi\left(\frac{L}{2}\right), \nonumber \\
	\theta_{R}(x) &=& \sum_{k\geq 0} \sqrt{\frac{2\pi g}{\lambda_k L}}
				\sin\left(\lambda_k x\right)
	i\left(a_{R,k}-a_{R,k}^{\dagger}\right)
	\nonumber \\ && +\theta(0),
	\label{eq:mode_exp_right}
\end{eqnarray}
where $\lambda_k = (2k+1)\pi/L$, and the $a_{\alpha,k}$ are canonical bosonic operators $[a_{\alpha,k},a^{\dagger}_{\beta,q}]=\delta_{\alpha,\beta}\delta_{k,q}$. Then, the Hamiltonian in each sector reduces to free bosonic modes
\begin{equation}
	\mathcal{H}_{0,L/R} = \sum_{k\geq 0}\epsilon_k (a^{\dagger}_{L/R,k} a_{L/R,k} +1/2),
\end{equation}
with the spectrum $\epsilon_k = v_F \lambda_k /g$. In addition it also supports a zero mode $\alpha_{L/R}$. We write explicitly
\begin{eqnarray}
	\alpha_{L} &=& e^{\frac{i}{2}\left(\pi\hat{m}+
				\pi \hat{n}_{-}\right)}
					\otimes I ,\nonumber \\
	\alpha_{R} &=& e^{\frac{i}{2}\left(\pi\hat{m}+
				\pi \hat{n}_{+}\right)}
	\otimes A,
\end{eqnarray}
where $A$ operates in the impurity-spin space alone and flips the spin. For spin-$1/2$ $A=S^x$, and the generalization to larger spins is immediate. The asymmetry between the operators $\alpha_L$ and $\alpha_R$ originates from the commutation relation in Eq.~(\ref{eq:phi_theta_comm}), which gives a nontrivial commutation relation between $\hat{m}$ and $\hat{n}_+$ but a trivial one between $\hat{m}$ and $\hat{n}_-$. One can directly verify that these operators commute with the Hamiltonian in the infinite backscattering limit, and are therefore zero modes. As the operators satisfy the commutation relation
\begin{equation}
	\alpha_L \alpha_R = e^{i\frac{\pi}{2}}\alpha_R\alpha_L,
\end{equation}
and have the property that $\alpha_{L,R}^4 = 1$, they are $\mathbb{Z}_4$ parafermions. The tunneling between the different states in the ground-state manifold is described by the operator $\hat{F} = e^{i(\pi-\Phi)/4}\alpha^{\dagger}_R\alpha_L$, which induces the transformation
\begin{eqnarray}
	\hat{F}\hat{m}\hat{F}^{\dagger} &=& \hat{m} + 1, \nonumber \\
	\hat{F}S_z\hat{F}^{\dagger} &=& -S_z,
\end{eqnarray}
while $\hat{n}_{\pm}$ remain unchanged by the transformation. The $\pi/4$-phase ensures that under time reversal $\hat{F}$ transforms as $T\hat{F}(\Phi)T^{-1} = \hat{F}(-\Phi)$.

This tunneling operator carries a fractional $e/2$ charge across the junction, as the charge density is the derivative of the $\theta$-field. To see this explicitly, one can examine the charge to the left and right of the impurity by defining
\begin{eqnarray}
	Q_{L} &=& \frac{e}{\pi}\int_{-\infty}^{0} dx\partial_x \theta(x) = 
				\frac{e}{\pi}[\theta(0)-\theta(-\infty)], \nonumber \\
	Q_{R} &=& \frac{e}{\pi}\int_{0}^{\infty} dx	\partial_x \theta(x) = 
				\frac{e}{\pi}[\theta(\infty)-\theta(0)],
\end{eqnarray}
and a direct calculation gives
\begin{eqnarray}
	\hat{F} Q_{L} \hat{F}^{\dagger} &=& Q_{L} - \frac{e}{2},
	\nonumber \\
	\hat{F} Q_{R} \hat{F}^{\dagger} &=& Q_{R} + \frac{e}{2}.
\end{eqnarray}

It should also be noted that $\hat{F}$ can be written using the original electronic degrees of freedom, using the identity $\partial_x\varphi(x) = -\pi\sigma^z(x)$, which recasts it as
\begin{eqnarray}
	\hat{F} &=& ie^{-i\Phi/4}
		e^{-i\frac{\pi}{2}
		\left(\hat{n}_+-\hat{n}_-\right)}\otimes A \nonumber \\ &&
		= i\exp\
		\left[i\frac{\pi}{2}\int_{-L/2}^{L/2}\!dx\sigma^z(x)
		\right]\otimes A,
		\label{eq:F_fermionic}
\end{eqnarray}
with $\sigma^z(x) = :\!\psi^{\dagger}_{\uparrow}(x)\psi_{\uparrow}(x)-\psi^{\dagger}_{\downarrow}(x)\psi_{\downarrow}(x)\!:$. This form emphasizes the many-body nature of the parafermionic state. We elaborate on the many-body fermionic picture in App.~\ref{app:elec_cut_wire}.

\section{Relaxing the strong backscattering limit}
\label{sec:relaxing_J}
For infinite $J_B$ the value of $\theta(0)$ is pinned to one of the four values of the ground state and the four-fold degeneracy is perfect. Relaxing this condition will allow finite coupling between the different states in the ground-state manifold, lifting some of the degeneracies. The strong backscattering induces a potential barrier for the $\theta(0)$ field (see Fig.~\ref{fig:low_e}), and the coupling between the states within the ground state manifold arises from two separate and independent contributions. The first one is tunneling between adjacent minima through the barrier, which can be calculated nonperturbatively by instanton methods. We label the amplitude of such tunneling events by $t_{\pi/2}$. The second contribution is tunneling between next-adjacent minima, whose $\theta(0)$ values are separated by $\pi$, and we label its amplitude by $t_{\pi}$. For simplicity, we omit from the Hamiltonian the less-relevant forward scattering term (see Sec.~\ref{sec:RG}).

Tunneling between adjacent minima requires the flipping of the spin, while tunneling between next-adjacent minima does not. Therefore, $t_{\pi/2}$ will be the dominant contribution when the exchange couplings are such that the impurity spin tends to follow the configuration of the edge electrons encoded in $\theta(0)$, while $t_{\pi}$ will be the dominant one when the spin tends to remain fixed, allowing only for single electrons to tunnel through without flipping. In the general case both of them will coexist in the effective low-energy Hamiltonian.

In order to evaluate $t_{\pi/2}$, we start from the backscattering Hamiltonian $\mathcal{H}_{B}$ of Eq.~(\ref{eq:H_bos_before_RG}), and write $\theta_{0}$ defined in Eq.~(\ref{eq:theta_0_def}) explicitly as
\begin{equation}
	\tan(4\theta_0) = \frac{2{\bf  J}_x\cdot{\bf  J}_y}
	{|{\bf  J}_x|^2-|{\bf  J}_y|^2},
\end{equation}
and the maximal energies $\pm B$ are given by
\begin{eqnarray}
	B^2 &=& \frac{{\bf  J}_x^2\!+\!{\bf  J}_y^2\!+\!
	\sqrt{\left({\bf  J}_x^2\!-\!{\bf  J}_y^2\right)^2\!+\!
	4({\bf  J}_x\cdot{\bf  J}_y)^2}}{2(\pi a)^2}
	\nonumber \\ &=& \frac{{\bf  J}_x^2\!+\!{\bf  J}_y^2}
	{2(\pi a)^2}\left( 1+
	\frac{J_B}{\sqrt{{\bf  J}_x^2\!+\!{\bf  J}_y^2}}
	\right).
\end{eqnarray}
For spin-$1/2$, for each value of $\theta(0)$ there are two eigenvalues of $\mathcal{H}_B$, given in Eq.~(\ref{eq:B_of_theta}), corresponding to opposite orientations of the impurity spin. Assuming ${\bf J}_x \times {\bf J}_y\neq 0$, the two levels are separated for every $\theta(0)$ (there are no degeneracies), and the height of the potential barrier separating adjacent minima is
\begin{eqnarray}
	\Delta B &=&
	\frac{\left({\bf J}_x^2\!+\!{\bf J}_y^2\right)^{\tfrac{1}{4}}}
	{\pi a} \nonumber \\ && \times
	\frac{\sqrt{\sqrt{{\bf  J}_x^2\!+\!{\bf  J}_y^2}+
	J_B}-
	\sqrt{\sqrt{{\bf  J}_x^2\!+\!{\bf  J}_y^2}-
	J_B}
	}{\sqrt{2}} \nonumber \\ &&
	\sim \frac{J_B}{\sqrt{2}\pi a}.
\end{eqnarray}

The role of the kinetic term that drives $\theta(0)$ is taken by a term that does not commute with $\theta(0)$. The first natural candidate for such a term is the relevant superconducting pairing at $x>L/2$, expressed in terms of the $\varphi$ field, which has non-trivial commutation relations with $\theta(0)$. However, the superconducting pairing is of the form $\sin[2\varphi(x)]$, which can only induce a $2\pi$ shift of $\theta(0)$, as a Cooper pair tunnels across the junction. Therefore it commutes with $\mathcal{H}_B$ and cannot give rise to tunneling between different minima.

We turn then to consider the kinetic term $v_F/(2\pi) \int\! dx (\partial_x\varphi)^2$. This term has non trivial commutation relations with $\theta(0)$ as $\left[\partial_x \varphi(x),\theta(0)\right] = i\pi\delta(x)$, subject to the regularization $\delta(0) = 1/a$, with $a$ the short-distance cutoff.
Keeping from the integral in the kinetic term only the derivative at $x=0$, we are left with $v_F a/(2\pi)\left[\partial_x\varphi(0)\right]^2$, and we define our pseudo-kinetic energy, driving $\theta(0)$, as
\begin{equation}
	\frac{p_\theta^2}{2m} = \frac{\pi v_F}{2a}
	\left[\frac{a}{\pi}\partial_x\varphi(0)\right]^2.
\end{equation}

\begin{figure}[t]
\includegraphics[width=0.45\textwidth]{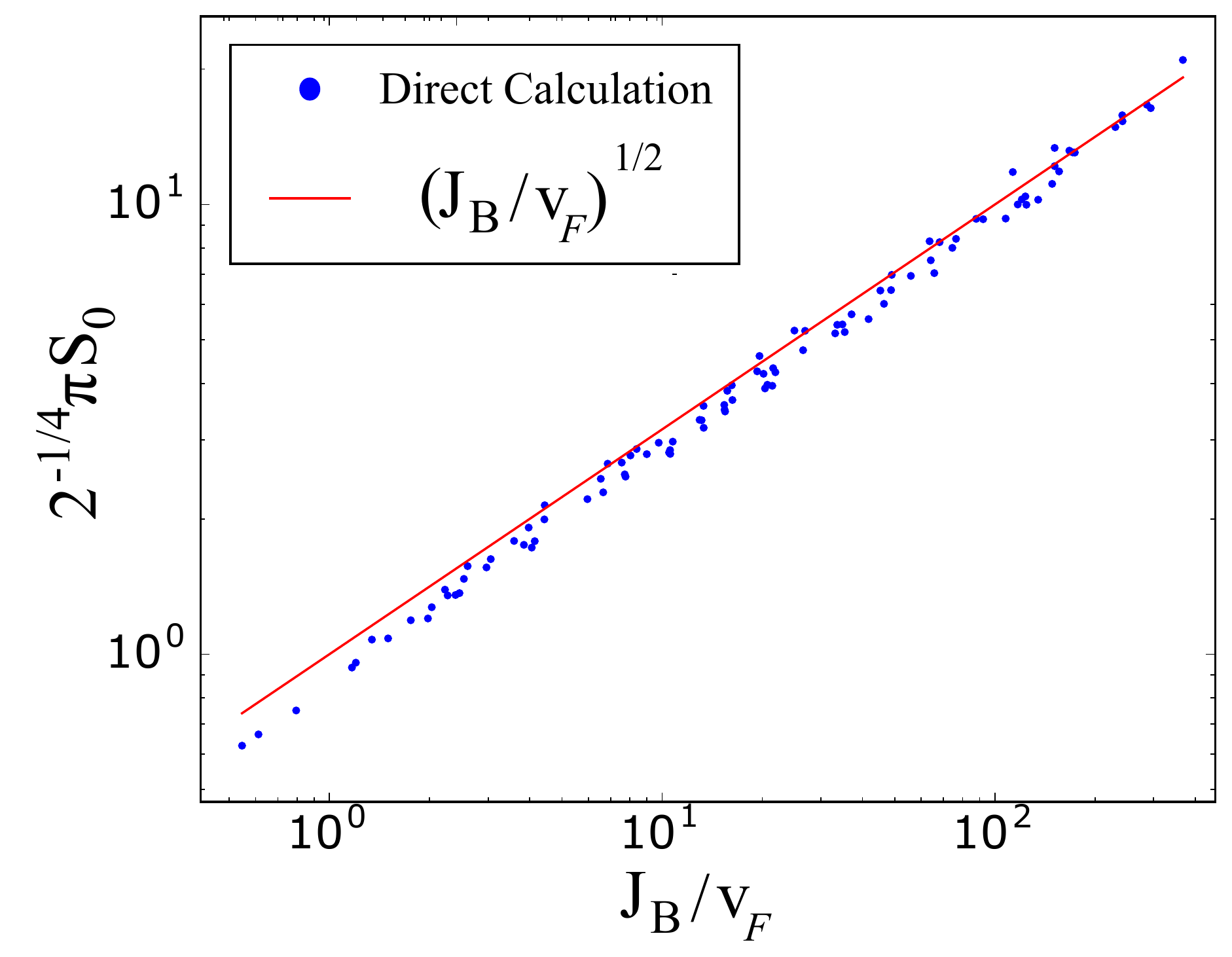}
% \vspace{0pt}
\caption{(color online) Numerical calculation of the classical instanton action $S_0$, for different randomly chosen values of ${\bf J}_x$ and ${\bf J}_y$. In these calculations the full dynamics of the impurity spin were considered, and the adiabatic limit was not taken explicitly. The red line shows a perfect square root behavior, while the blue dots are the results of the numerical calculations.}
\label{fig:S_0}
\end{figure}

The contribution of a single instanton to the matrix element between two adjacent states, and therefore to $t_{\pi/2}$, is proportional to $\exp(-S_0)$ with $S_0$ the action of the classical path starting at $|\theta_0,S^z\rangle$ and ending at $|\theta_0+\pi/2,-S^z\rangle$, under the inverted potential $-\mathcal{H}_B$. In order to evaluate the action in analytical form, we first assume that the dynamics of the spin is much faster than the dynamics of the local field $\theta(0)$. Under this assumption we can take the adiabatic limit in which the spin follows the effective magnetic field 
\begin{equation}
	{\bf B}[\theta(0)] = \frac{{\bf J}_x}{\pi a} \sin[2\theta(0)]
	-\frac{{\bf J}_y}{\pi a} \cos[2\theta(0)]
	\label{eq:eff_magnetic_field}
\end{equation}
at each point [see Fig.~\ref{fig:low_e}(b).] We have thus mapped the problem to a tunneling problem of $\theta(0)$ between minima of the periodic potential $V[\theta(0)]$ given by
\begin{eqnarray}
	V(x) &=& -\frac{1}{\pi a}
	\sqrt{\frac{{\bf J}_x^2+{\bf J}_y^2}{2}}
	\times \nonumber \\ &&
	\sqrt{1-
	\frac{{\bf J}_x^2-{\bf J}_y^2}{{\bf J}_x^2+{\bf J}_y^2}
	\cos(4x)-
	\frac{2{\bf J}_x\cdot{\bf J}_y}{{\bf J}_x^2+{\bf J}_y^2}
	\sin(4x)}, \nonumber \\ &&
\end{eqnarray}
which can be seen in Fig.~\ref{fig:low_e}(a). The action along the classical path in the adiabatic limit is given by
\begin{eqnarray}
	S_0 &\simeq & \sqrt{\frac{a}{\pi v_F}}
		\int_{\theta_0}^{\theta_0+\pi/2}\! d\theta 
		\sqrt{V(\theta)+B} 
		\nonumber \\ &\simeq &
		\frac{2^{1/4}}{\pi}\sqrt{\frac{J_B}{v_F}}.
\end{eqnarray}
We have also calculated the action numerically, taking into account the full dynamics of the impurity spin, and verified that the action scales with $\sqrt{J_B/v_F}$ even when the adiabatic limit is not taken explicitly (see Fig.~\ref{fig:S_0}).

The exponential dependence is the dominant scaling of the tunneling amplitude
\begin{equation}
	t_{\pi/2} \sim e^{-\frac{2^{1/4}}{\pi}\sqrt{\frac{J_B}{v_F}}},
\end{equation}
and we will not evaluate the prefactor to $t_{\pi/2}$. As $t_{\pi/2}$ is exponentially small in $\sqrt{J_B/v_F}$, for strong backscattering the degeneracy lifting is small. This parameter can be increased by working with spins larger than $1/2$, as the height of the potential barrier between the wells depends linearly on the size of the spin.

In the calculation above we omitted the forward scattering term $\mathcal{H}_F = -{\bf J}_z\cdot{\bf S} \partial_x\varphi(0)/\pi$. In the instanton language formulated here, this term is equivalent to a term linear in $p_\theta$ which is coupled to the local impurity spin. While a term solely linear in $p_\theta$ will not affect the calculations, as it can be effectively incorporated into the quadratic term by a constant shift in the definition of $p_{\theta}$, the coupling to the impurity spin degrees of freedom means that it will also affect its dynamics. Adding this term changes the classical equations of motions for the spin degrees of freedom to be
\begin{equation}
	\dot{\bf S} = \frac{1}{\pi a}\left[{\bf J}_x \sin[2\theta(0)]-
	{\bf J_y}\cos[2\theta(0)]-\pi p_{\theta}{\bf J}_z
	\right]\times {\bf S}.
\end{equation}
The magnitude of $p_{\theta}$ can be evaluated by $\sim \sqrt[4]{J_B/v_F}$, therefore the added term can be neglected as long as $(J_B/v_F)^{3/4} \gg |{\bf J}_z|/v_F$. Since $J_B$ is relevant, while ${\bf J}_z$ is marginal, this assumption is justified, and we may omit this term altogether.

We turn to evaluate $t_{\pi}$, which correspond to tunneling between next-adjacent minima. This type of process does not require the spin to flip, and $t_{\pi}$ will be dominant when the spin does not easily follow the instantaneous configuration of the electrons. One can therefore fix the spin orientation and solve the $1D$ problem of tunneling through a barrier, a task which is done in App.~\ref{app:elec_cut_wire}, and which yields the scaling
\begin{equation}
	t_{\pi} \sim e^{-\frac{J_B}{v_F}}.
\end{equation}

The low-energy subspace is separated from the rest of the spectrum by three distinct gaps:
\begin{enumerate}
\item A gap of size $\sim\! B$ separating it from excitations associated with the magnetic orientation of the impurity spin at $x=0$, fixing the orientation of the impurity spin and of the edge electrons.

\item A gap of size $\sim\!\Delta $ separating it from quasi-particle excitations in the superconducting leads.

\item A gap of size $\sim\!\delta/g = \pi v_F/gL$ separating it from excitations of the free modes inside the junction itself.

\end{enumerate}
In order for the parafermions to remain protected even for finite $J_B$, we require that $t_{\pi/2}$ and $t_{\pi}$ will be much smaller than these gaps. This is achieved in the limit $J_B/v_F \gg 1$.

Assuming that the system resides in the regime where the ground state manifold is well separated from the rest of the spectrum, the Hamiltonian within the ground-state manifold can be written using the $\hat{F}$ operator that tunnels between adjacent states within the subspace, as
\begin{eqnarray}
	\mathcal{H}_{gs} = 
	t_{\pi/2}\hat{F}(\Phi) + 
	t_{\pi}\hat{F}^2(\Phi)+
	{\rm h.c.}.
	\label{eq:eff_H_gs_phi}
\end{eqnarray}
In Fig.~\ref{fig:eff_H_energies}(a) we plot the energy dependence on the superconducting phase difference $\Phi$ across the gap, and reveal that the Josephson current is $8\pi$-periodic in the phase. The Hamiltonian of Eq.~(\ref{eq:eff_H_gs_phi}) is identical to the one derived in Ref.~\cite{Zhang_2014}, where a superconductor-quantum spin Hall edge-superconductor junction was studied, but with a strong pair-backscattering term in the junction, which respected time reversal symmetry. The energy dependence in Fig.~\ref{fig:eff_H_energies} also matches the results in Refs.~\cite{Peng_2016} and ~\cite{Hui_2017}, which considered the weak-coupling regime.

\section{Summary and Outlook}
In this work we have studied a basic setup consisting of an interacting quantum spin Hall edge Josephson junction coupled to an impurity spin. We focused on the conditions under which stable $\mathbb{Z}_4$ parafermions are created in the setup, and the characteristics of these exotic quasiparticles. In order for the parafermions to be well defined and stable the exchange coupling between the electrons in the edge and the impurity spin has to be strongly anisotropic, and some repulsive interaction between the electrons in the junction must exist. Under these conditions, the renormalization group flow of the setup leads to a regime dominated by the backscattering of the edge electrons off the impurity spin, and the Hamiltonian describing the impurity spin and the edge electrons coupled to it has an approximately four-fold degenerate ground state. The four-fold degeneracy can be seen as a combination of the two-fold ground state degeneracy in topological Josephson junctions due to the existence of Majorana fermions, with the two possible magnetic orientations of the spin impurity.

The four-fold degeneracy is related to the time reversal symmetry of the system, and a basis can be chosen in which the states that constitute the ground state manifold are related to each other by time reversal transformation. In this basis, the impurity spin in each state is in one of two possible orientations, and a pair of states with identical orientation is distinguished by the fermion parity along the edge.

The ground state manifold is described by $\mathbb{Z}_4$ parafermions, which are located to the left and to the right of the impurity spin, and allow for tunneling of fractional charge $e/2$ between the superconductors. For infinite backscattering $J_B$, the parafermions are decoupled and the degeneracy is perfect. Finite values of $J_B$ lift the degeneracies as the parafermions are weakly coupled with strength exponentially small in $\sqrt{J_B/v_F}$. The resulting low-energy spectrum is $8\pi$ periodic as a function of the phase bias across the junction $\Phi$.

An interesting feature of the parafermions presented here is the fact they are not composed solely of the electrons in the quantum spin Hall edge, but also involve the spin impurity. We believe that this may provide a unique ability to create and manipulate these parafermions. In a proposed setup, the spin impurity will be an effective description of a quantum dot placed in proximity to the edge, that can be controlled and measured. While $\mathbb{Z}_n$ parafermions are considered as possible building-blocks for implementing universal quantum computation~\cite{Nayak_2008}, this is not the case for even values of $n$, which only allows a limited set of computational operations~\cite{Hutter_2016}.

The system presented here, and its suggested experimental realization, can serve as a starting point for further research. Of particular interest are calculations of observables in experimentally relevant setups, such as electric conductance and noise measurements, the response of the system to {\em AC} drives and the potential signatures of the exotic parafermionic states in the quantum impurity. From a theoretical point of view, a numerical analysis of the flow to the different phases beyond weak coupling could be of value.

\begin{acknowledgments}
The authors would like to thank A. Bruch, C. Karrasch, D. Litinski and Y. Peng for helpful discussions. We acknowledge financial support by  the  Deutsche  Forschungsgemeinschaft  (CRC 183 under project C02) and by the Minerva foundation.
\end{acknowledgments}

\appendix

\section{Perturbative Renormalization Group and Poor Man's Scaling Calculations}
\label{app:RG}
In this section we will present the derivation of the renormalization group flow equations in Eqs.~(\ref{eq:RG_bosonic_leading_order}) and ~(\ref{eq:RG_bosonic_second_order}). We will then qualitatively corroborate our results by comparing them with a Poor Man's scaling done on the noninteracting ($g=1$) model.

The action for the Hamiltonian of Eq.~(\ref{eq:H_bos_before_RG}) can be written as $S = S_0 + S_{\Delta} + S_{B} + S_{F}$ with
\begin{eqnarray}
	S_0 &=& \frac{1}{2\pi g} \int\!dx dt
	u_F\left[\partial_x \theta(x,t)\right]^2+
	\frac{1}{u_F}\left[\partial_t \theta(x,t)\right]^2,
	\nonumber \\
	S_{\Delta} &=& \frac{\Delta}{\pi a}
	\int\!dx dt \Theta\left(|x|-\frac{L}{2}\right)
	\sin\left[2\varphi(x,t)-\Phi(x)\right], \nonumber \\
	S_{B} &=& \frac{1}{\pi a}\int\!dt \left\{
	\cos\left[2\theta(0,t)\right]{\bf  J}_y-
	\sin\left[2\theta(0,t)\right]{\bf  J}_x\right\}\cdot {\bf  S},
	\nonumber \\
	S_{F} &=& -\frac{1}{\pi}
			\int\!dt \partial_x\varphi(0,t) {\bf  J}_z \cdot {\bf  S},
\end{eqnarray}
where $u_F = v_F/g$, $\varphi(x,t)$ is the dual field to $\theta(x,t)$ and the $9$ exchange coefficients are represented by the three real vectors $\vec{J}_i$, which have dimensions of energy times length.

It is convenient to expand $\theta(x,t)$ and $\varphi(x,t)$ in the bosonic modes
\begin{eqnarray}
	\theta(x,t) &=& i\sqrt{g} \sum_{q>0} \sqrt{\frac{\pi}{2qV}}
	\big[e^{iq(x-u_F t)}d_{q,\uparrow} + e^{iq(x+u_F t)}
	d^{\dagger}_{q,\downarrow} \nonumber \\ && - {\rm h.c.}\big] +
	\frac{\hat{\theta}_{\uparrow}-\hat{\theta}_{\uparrow}}{2},
	\nonumber \\
	\varphi(x,t) &=& \frac{i}{\sqrt{g}} \sum_{q>0}
	\sqrt{\frac{\pi}{2qV}}
	\big[e^{iq(x-u_F t)}d_{q,\uparrow} - e^{iq(x+u_F t)}
	d^{\dagger}_{q,\downarrow} \nonumber \\ && - {\rm h.c.}\big] +
	\frac{\hat{\theta}_{\uparrow}+\hat{\theta}_{\uparrow}}{2},
	\label{app_eq:operator_field_exp}
\end{eqnarray}
where $d_{q,\sigma}$ are canonical bosonic annihilation and creation operators, $\hat{\theta}_{\sigma}$ are phase factors with $[\hat{\theta}_{\uparrow},\hat{\theta}_{\downarrow}]=-i\pi$ and $V$ is the size of the system. An important relation is
\begin{equation}
	\frac{\partial\theta(x,t)}{\partial x} = 
	-\frac{g}{u_F}\frac{\partial\varphi(x,t)}{\partial t}
	\;,\;\; 
	g\frac{\partial\varphi(x,t)}{\partial x} = 
	-\frac{1}{u_F}\frac{\partial\theta(x,t)}{\partial t}.
	\label{app_eq:deriv_fields_relation}
\end{equation}

For our renormalization group process, we will enforce a hard-cutoff $\Lambda/u_F$ and divide our modes to fast ones, lying in the interval $\Lambda ' < q \leq \Lambda$, for some $\Lambda ' = \Lambda/(1+dl)$ and slow ones $q \leq \Lambda '$. We start by calculating an effective ${S'}_{\rm eff}$ by averaging on the fast modes
\begin{equation}
	{S'}_{\rm eff} = -\ln \langle e^{-S} \rangle_{>},
\end{equation}
and then rescaling $dx \to (1+dl) dx$ and $dt \to (1+dl)dt$ to restore the original cutoff. The different orders in the renormalization group flow equations correspond to the cumulant expansion of ${S'}_{\rm eff}$.

\subsection{Leading order}
The leading order in the cumulant expansion is calculated by taking the average on the action with respect to the fast modes ${S'}_{\rm eff}^{(1)} = \langle S \rangle_>$. Using the identity
\begin{equation}
	\langle e^{A} \rangle =
		e^{\langle A \rangle +
		\frac{1}{2}\left(\langle A^2\rangle - \langle A \rangle ^2\right)},
\end{equation}
which is true for any operator $A$ linear in creation and annihilation operators, we have
\begin{eqnarray}
	\langle e^{\pm 2i\theta(x,t)} \rangle _> &=& 
	e^{\pm 2i\theta_<(x,t)-2g\sum_{q>\Lambda '}^{\Lambda}
	\frac{\pi}{qV}}, \nonumber \\ 
	\langle e^{\pm 2i\varphi(x,t)} \rangle _> &=& 
	e^{\pm 2i\varphi_<(x,t)-\frac{2}{g}\sum_{q>\Lambda '}^{\Lambda}
	\frac{\pi}{qV}}. 
\end{eqnarray}
Taking the continuum limit $\sum_{q}2\pi/V \to \int\! dq$ we arrive at
\begin{eqnarray}
	\langle \cos[2\theta(0,t)] \rangle _> &=&
	\cos[2\theta_<(0,t)](1+dl)^{-g}, \nonumber \\
	\langle \sin[2\theta(0,t)] \rangle _> &=&
	\sin[2\theta_<(0,t)](1+dl)^{-g}, \nonumber \\
	\langle \sin[2\varphi(0,t)] \rangle _> &=&
	\sin[2\theta_<(0,t)](1+dl)^{-\frac{1}{g}},
\end{eqnarray}
and the average over $\partial_x \varphi(0,t)$ does not contribute to
\begin{equation}
	\langle \frac{\partial\varphi(0,t)}{\partial x}\rangle_> = 
	\frac{\partial\varphi_<(0,t)}{\partial x}.
\end{equation}

If one chooses to stop at the leading order, the process will be complete by the rescaling back to the original cutoff. The superconductor pairing is multiplied by $(1+dl)^2$ while the local exchange interactions are only multiplied by $(1+dl)$, leading to the action terms
\begin{eqnarray}
	{S'}_\Delta &=& (1+dl)^{2-\frac{1}{g}}S_{\Delta},
	\nonumber \\ 
	{S'}_{B} &=& (1+dl)^{1-g}S_{B}, \nonumber \\
	{S'}_{F} &=& S_{F},
\end{eqnarray}
where the rescaling of ${S'}_{F}$ is zero due to the spatial derivative.

\subsection{Second order in the exchange interactions}
Seeing that $g=1$ is a fixed point for the local exchange interaction at leading order, we now turn to calculate the second order in the cumulant expansion, while restricting ourselves to the exchange interaction
\begin{equation}
	{S'}^{(2)}_{\rm eff} = -\frac{1}{2}\left[\langle (S_{B}+S_{F})^2\rangle_>-
	\langle S_{B}+S_{F}\rangle^2_>
	\right].
	\label{app_eq:2nd_order}
\end{equation}
Omitting the $S_\Delta$ contribution will have little effect, as these parts of the action acts in different points in space, thus the cross-correlation between them are expected to be small. We shall restrict ourselves to the case of spin-$1/2$ in this calculation.

\subsubsection{Some useful auxiliary results}
The second order expansion is more complicated, and it is convenient to have in advance some useful auxiliary results, pertaining to the commutation relations between the fields and the normal ordering operators.

Using the operators expansions in Eq.~(\ref{app_eq:operator_field_exp}) we can write the commutation relations between the fields as
\begin{eqnarray}
	\left[\theta(x_1,t_1),\theta(x_2,t_2)\right] &=& 
	-ig\sum_{q<\Lambda} \frac{2\pi}{qV}
	\cos(q\Delta x)\sin(qu_F\Delta t)
	, \nonumber \\
	\left[\varphi(x_1,t_1),\varphi(x_2,t_2)\right] &=&
	-\frac{i}{g}\sum_{q<\Lambda} \frac{2\pi}{qV}
	\cos(q\Delta x)\sin(qu_F\Delta t)
	, \nonumber \\ 
	\left[\varphi(x_1,t_1),\theta(x_2,t_2)\right] &=&
	i\sum_{q<\Lambda} \frac{2\pi}{qV}
	\sin(q\Delta x)\cos(qu_F\Delta t) \nonumber \\ && + \frac{i\pi}{2},
\end{eqnarray}
with $\Delta x = x_1-x_2$ and $\Delta t = t_1-t_2$. In order to calculate the normal ordering we divide each field into a parts containing only creation ($+$) and annihilation ($-$) operators $$\theta(x,t) = \theta^+(x,t) + \theta^-(x,t)+(\hat{\theta}_{\uparrow}-\hat{\theta}_{\downarrow})/2$$ and $$\varphi(x,t) = \varphi^+(x,t) + \varphi^-(x,t)+(\hat{\theta}_{\uparrow}+\hat{\theta}_{\downarrow})/2.$$ Using the commutation relations between them
\begin{eqnarray}
	\left[\theta^-(x_1,t_1),\theta^+(x_2,t_2) \right] &=&
	g\sum_{q<\Lambda}\frac{\pi}{qV}\cos(q\Delta x)e^{-iqu_F\Delta t},
	\nonumber \\ 
	\left[\varphi^-(x_1,t_1),\varphi^+(x_2,t_2) \right] &=&
	\frac{1}{g}
	\sum_{q<\Lambda}\frac{\pi}{qV}\cos(q\Delta x)e^{-iqu_F\Delta t},
	\nonumber \\
\end{eqnarray}
\begin{widetext}
\noindent we can normal order exponential operators of $\theta$ as
\begin{eqnarray}
	:\!e^{in\theta(x,t)}\!: &=& e^{in\theta(x,t)}
			e^{\frac{n^2}{2}g
				\sum_{q}\frac{\pi}{qV}
						}, \nonumber \\
	:\!e^{in\theta(x_1,t_1)+im\theta(x_2,t_2)}\!: &=& 
	e^{in\theta(x_1,t_1)+im\theta(x_2,t_2)}
	e^{\frac{n^2+m^2}{2}g\sum_q \frac{\pi}{qV}}
	e^{nmg\sum_{q}\frac{\pi}{qV}\cos(q\Delta x)\cos(qu_F\Delta t)}
\end{eqnarray}
and further write the multiplication rules for normal-ordered exponential operators that will come about during the calculation of $\langle S_{B}^2 \rangle$
\begin{eqnarray}
	:\!e^{in\theta(x_1,t_1)}\!::\!e^{im\theta(x_2,t_2)}\!: &=& 
	:\!e^{in\theta(x_1,t_1)+im\theta(x_2,t_2)}\!:
	e^{nmg\sum_{q}\frac{\pi}{qV}\cos(q\Delta x)
	e^{-iqu_F\Delta t}},
	\nonumber \\ 
	2:\!\sin[n\theta(x_1,t_1)]\!::\!\cos[m\theta(x_2,t_2)]\!: &=&
	:\!\sin[n\theta(x_1,t_1)+m\theta(x_2,t_2)]\!:
	e^{nmg\sum_{q}\frac{\pi}{qV}\cos(q\Delta x)
	e^{-iqu_F\Delta t}} + \nonumber \\ &&
	:\!\sin[n\theta(x_1,t_1)-m\theta(x_2,t_2)]\!:
	e^{-nmg\sum_{q}\frac{\pi}{qV}\cos(q\Delta x)
	e^{-iqu_F\Delta t}}, \nonumber \\
	2:\!\cos[n\theta(x_1,t_1)]\!::\!\sin[m\theta(x_2,t_2)]\!: &=&
	:\!\sin[n\theta(x_1,t_1)+m\theta(x_2,t_2)]\!:
	e^{nmg\sum_{q}\frac{\pi}{qV}\cos(q\Delta x)
	e^{-iqu_F\Delta t}} - \nonumber \\ &&
	:\!\sin[n\theta(x_1,t_1)-m\theta(x_2,t_2)]\!:
	e^{-nmg\sum_{q}\frac{\pi}{qV}\cos(q\Delta x)
	e^{-iqu_F\Delta t}}, \nonumber \\
	2:\!\sin[n\theta(x_1,t_1)]\!::\!\sin[m\theta(x_2,t_2)]\!: &=&
	:\!\cos[n\theta(x_1,t_1)-m\theta(x_2,t_2)]\!:
	e^{-nmg\sum_{q}\frac{\pi}{qV}\cos(q\Delta x)
	e^{-iqu_F\Delta t}} - \nonumber \\ &&
	:\!\cos[n\theta(x_1,t_1)+m\theta(x_2,t_2)]\!:
	e^{nmg\sum_{q}\frac{\pi}{qV}\cos(q\Delta x)
	e^{-iqu_F\Delta t}}, \nonumber \\
	2:\!\cos[n\theta(x_1,t_1)]\!::\!\cos[m\theta(x_2,t_2)]\!: &=&
	:\!\cos[n\theta(x_1,t_1)-m\theta(x_2,t_2)]\!:
	e^{-nmg\sum_{q}\frac{\pi}{qV}\cos(q\Delta x)
	e^{-iqu_F\Delta t}} + \nonumber \\ &&
	:\!\cos[n\theta(x_1,t_1)+m\theta(x_2,t_2)]\!:
	e^{nmg\sum_{q}\frac{\pi}{qV}\cos(q\Delta x)
	e^{-iqu_F\Delta t}}.
	\label{app_eq:normal_order}
\end{eqnarray}
It is also useful to have at hand the commutation relations between the derivative of $\varphi$ and the exponential operators of $\theta$
\begin{eqnarray}
	\left[e^{in\theta(x_1,t_1)},\partial_{x_2} \varphi(x_2,t_2)
	\right] &=&
	n\sum_{q}\frac{2\pi}{V}\cos(q\Delta x)\cos(qu_F\Delta t)
	e^{in\theta(x_1,t_1)},
	\nonumber \\
	\left[\cos[n\theta(x_1,t_1)],\partial_{x_2}
	 \varphi(x_2,t_2)\right] &=&
	n\sum_{q}\frac{2\pi}{V}\cos(q\Delta x)\cos(qu_F\Delta t)
	i\sin[n\theta(x_1,t_1)], \nonumber \\
	\left[\sin[n\theta(x_1,t_1)],\partial_{x_2}
	\varphi(x_2,t_2)\right] &=&
	-n\sum_{q}\frac{2\pi}{V}\cos(q\Delta x)\cos(qu_F\Delta t)
	i\cos[n\theta(x_1,t_1)].
	\label{app_eq:comm_with_derphi}
\end{eqnarray}
\end{widetext}
Finally, we note that for $S=1/2$ the following identity holds
\begin{equation}
	\left({\bf A} \cdot {\bf S}\right)
	\left({\bf B} \cdot {\bf S}\right) = 
	i\left({\bf A}\times {\bf B}\right)\cdot {\bf S} + 
	{\bf A}\cdot{\bf B}.
	\label{app_eq:vec_iden}
\end{equation}

\subsubsection{Operators Product Expansion}
We now turn to calculate the operator product expansion (OPE) of the different terms appearing in the second order equations. We will treat separately each of the terms in Eq.~(\ref{app_eq:2nd_order}). , $\langle S_{F}^2 \rangle_> - \langle S_{F} \rangle_>^2$ and $\langle \left\{S_{B},S_{F}\right\} \rangle_> - \left\{\langle S_{B} \rangle_>,\langle S_{F} \rangle_>\right\}$.

We begin by examining $\langle S_{B}^2 \rangle_> - \langle S_{B} \rangle_>^2$. Writing it explicitly we have
\begin{widetext}
\begin{eqnarray}
	\langle S_{B}^2 \rangle_> - \langle S_{B} \rangle_>^2 &=&
	\frac{1}{(\pi a)^2}\int\! dt_1 dt_2 \langle 
	\cos[2\theta(0,t_1)]\cos[2\theta(0,t_2)]|{\bf  J}_y|^2 +
	\sin[2\theta(0,t_1)]\sin[2\theta(0,t_2)]|{\bf  J}_x|^2\rangle_>-
	\nonumber \\ &&
	\frac{1}{(\pi a)^2}\int\! dt_1dt_2 \langle 
	\cos[2\theta(0,t_1)]\rangle_>
	\langle \cos[2\theta(0,t_2)]\rangle_>
	|{\bf  J}_y|^2 - \nonumber \\ &&
		\frac{1}{(\pi a)^2}\int\! dt_1dt_2 \langle 
	\sin[2\theta(0,t_1)]\rangle_>\langle 
	\sin[2\theta(0,t_2)]\rangle_>
	|{\bf  J}_x|^2 + \nonumber \\ &&
	\frac{i}{(\pi a)^2}\int\! dt_1 dt_2 \langle 
	\cos[2\theta(0,t_1)]\sin[2\theta(0,t_2)]-
	\sin[2\theta(0,t_1)]\cos[2\theta(0,t_2)]\rangle_>
	\left({\bf  J}_x\times{\bf  J}_y\right)\cdot{\bf  S}-
	\nonumber \\ &&
	\frac{i}{(\pi a)^2}\int\! dt_1 dt_2 \left\{ \langle 
	\cos[2\theta(0,t_1)]\rangle_>\langle\sin[2\theta(0,t_2)]\rangle_>-
	\langle\sin[2\theta(0,t_1)]\rangle_>\langle\cos[2\theta(0,t_2)]
	\rangle_>\right\}
	\left({\bf  J}_x\times{\bf  J}_y\right)\cdot{\bf  S}
	- \nonumber \\ &&
	\frac{1}{(\pi a)^2}\int\! dt_1 dt_2 \langle 
	\cos[2\theta(0,t_1)]\sin[2\theta(0,t_2)]+
	\sin[2\theta(0,t_1)]\cos[2\theta(0,t_2)]\rangle_>
	{\bf  J}_x\cdot{\bf  J}_y + \nonumber \\ &&
	\frac{1}{(\pi a)^2}\int\! dt_1 dt_2
	\left\{\langle 
	\cos[2\theta(0,t_1)]\rangle_>\langle\sin[2\theta(0,t_2)]\rangle_>+
	\langle\sin[2\theta(0,t_1)]\rangle_>
	\langle\cos[2\theta(0,t_2)]\rangle_>\right\}
	{\bf  J}_x\cdot{\bf  J}_y,
\end{eqnarray}

where we have used Eq.~(\ref{app_eq:vec_iden}) to treat the different impurity spin operators. Repeated application of the identities in Eq.~(\ref{app_eq:normal_order}) allows us to recast the $|{\bf  J}_y|^2$ term as
\begin{eqnarray}
	&&\frac{|{\bf  J}_y|^2}{(\pi a)^2}\int\! dt_1 dt_2 \big\{
	\langle 
	\cos[2\theta(0,t_1)]\cos[2\theta(0,t_2)]\rangle_> -
	\langle 
	\cos[2\theta(0,t_1)]\rangle_>\langle\cos[2\theta(0,t_2)]\rangle_>
	\big\} = \nonumber \\ 
	&& \;\;\;\;\;\;\;\;\;\;\;\;\;\;\;\;\;\;\;\;\;\;\;\;
	\frac{|{\bf  J}_y|^2}{(\pi a)^2}gdl 
	\int\!dt_1 dt_2 e^{-i\Lambda u_F(t_1-t_2)}
	\bigg\{\!:\!\cos[2\theta_<(0,t_1)-2\theta_<(0,t_2)]\!:
	e^{-4g\sum\frac{\pi}{qV}\left[1-e^{-iqu_F(t_1-t_2)}\right]}-
	\nonumber \\ 
	&& \;\;\;\;\;\;\;\;\;\;\;\;\;\;\;\;\;\;\;\;\;\;\;\;
	:\!\cos[2\theta_<(0,t_1)+2\theta_<(0,t_2)]\!:
	e^{-4g\sum\frac{\pi}{qV}\left[1+e^{-iqu_F(t_1-t_2)}\right]}
	\bigg\}.
\end{eqnarray}
\end{widetext}
As the contributions to the renormalization group come from the area where $t_1$ and $t_2$ are close, we can change variables as $T=(t_1+t_2)/2$ and $\Delta t = t_1-t_2$ and expand for small $\Delta t$ to get
\begin{eqnarray}
	:\!\cos[2\theta_<(0,t_1)-2\theta_<(0,t_2)]\!: &\simeq & 
	\cos[2\Delta t \partial_T\theta_<(0,T)] 
	\nonumber \\ &\simeq & 
	1-[2(\Delta t)^2\partial_T\theta_<(0,T)]^2,
	\nonumber \\
	:\!\cos[2\theta_<(0,t_1)+2\theta_<(0,t_2)]\!: &\simeq & 
	\cos[\theta_<(0,T)],
\end{eqnarray}
and integrating over $\Delta t$ in the range $u_F|\Delta t| \leq \pi/2\Lambda$ will give us the new contributions to the action. The first term will give rise to a local operator of the form $(\partial_t\theta)^2$ which is a forward potential scattering at the origin, and the second term will generate a local pair-scattering $\cos[4\theta(t)]$. Both of these terms are irrelevant near $g=1$, and we can ignore them. It is straightforward to convince oneself that the $|{\bf  J}_y|^2$ and the ${\bf  J}_x\cdot{\bf  J}_y$ terms will similarly contribute only operators that are irrelevant near $g=1$ and can be ignored.

We are therefore left with the $({\bf  J}_x\times{\bf  J}_y)\cdot {\bf  S}$ term. To calculate its contribution we repeat the above process, writing
\begin{widetext}
\begin{eqnarray}
	&& \frac{i}{(\pi a)^2}
	\int\!dt_1dt_2 \bigg\{
	\langle \big[
	\cos[2\theta(0,t_1)],\sin[2\theta(0,t_2)]\big]\rangle_> -
	\big[\langle 
	\cos[2\theta(0,t_1)]\rangle_>,\langle\sin[2\theta(0,t_2)]\rangle_>
	\big]\bigg\} =
	\nonumber \\ 
	&& \;\;\;\;\;\;\;\;\;\;\;\;\;\;\;\;\;\;\;\;\;\;\;\;
	-\frac{2i}{(\pi a)^2}gdl
	\int\!dt_1dt_2 e^{-i\Lambda u_F (t_1-t_2)-
	2g\sum\frac{2\pi}{qV}\left[1-e^{-iqu_F(t_1-t_2)}\right]
	}
	:\!\sin[2\theta_<(0,t_1)-2\theta_<(0,t_2)]\!:,
\end{eqnarray}
\end{widetext}
and again we are interested in the region where $t_1$ and $t_2$ are close, leading us to the coordinate substitution to $T$ and $\Delta t$. Expanding $\sin[2\theta_<(0,t_1)-2\theta_<(0,t_2)]\simeq 2\Delta t\partial_T\theta_<(0,T)$ we arrive at the term
\begin{eqnarray}
	&& -\frac{8dl}{(\pi a \Lambda u_F)^2}
	{\rm Im}\left\{
	\int_0^{\pi/2}\! dx  x e^{ix-2gJ(x)}
	\right\} \times \nonumber \\ && 
	\int\!dTg\partial_T\theta_<(0,T)\left({\bf  J}_x\times
	{\bf  J}_y
	\right)\cdot {\bf  S},
\end{eqnarray}
with
\begin{equation}
	J(x) = \int_0^{1}\!ds\frac{1-e^{ixs}}{s}.
\end{equation}
We finally use the relation from Eq.~(\ref{app_eq:deriv_fields_relation}) to substitute the time derivative of $\theta$ with the position derivative of $\varphi$ to connect to the $\mathcal{S}_{FS}$, and we get that to second order the renormalization group equation for ${\bf  J}_z$ is
\begin{eqnarray}
	\frac{d{\bf  J}_z}{dl} &=& 
	\frac{4\pi g^3}{(\pi a \Lambda)^2}
	{\rm Im}\left\{
	\int_0^{\pi/2}\! dx  x e^{ix-2gJ(x)}
	\right\} \times \nonumber \\ && 
	\frac{1}{v_F}\left({\bf  J}_x\times
	{\bf  J}_y
	\right)\cdot {\bf  S}.
\end{eqnarray}
Numerical analysis of the prefactor shows that it is positive for values of $g$ smaller than $g_c \simeq 1.27$.

We now turn to calculate the mixed terms of $S_{F}$ and $S_{B}$, which can be written using Eq.~(\ref{app_eq:vec_iden}) as
\begin{widetext}
\begin{eqnarray}
	\langle \left\{S_{B},S_{F}\right\}\rangle_>\!-\!
	\left\{\langle S_{B}\rangle_>,\langle S_{F}\rangle_>\right\}
	&=& -\frac{i}{\pi a}\int\! dt_1 dt_2
	\bigg\{
	\langle \cos[2\theta(0,t_1)]\partial_x\varphi(0,t_2)
	-\partial_x\varphi(0,t_2)\cos[2\theta(0,t_1)]
	\rangle_>-\nonumber \\ &&
	\langle \cos[2\theta(0,t_1)]\rangle_>
	\langle\partial_x\varphi(0,t_2)\rangle_>
	+\langle\partial_x\varphi(0,t_2)\rangle_>
	\langle\cos[2\theta(0,t_1)]\rangle_>
	\bigg\}\left({\bf  J}_y\times{\bf  J}_z\right)\cdot{\bf  S}
	\nonumber \\ &&
	-\frac{1}{\pi a}\int\! dt_1 dt_2
	\bigg\{
	\langle \cos[2\theta(0,t_1)]\partial_x\varphi(0,t_2)
	+\partial_x\varphi(0,t_2)\cos[2\theta(0,t_1)]
	\rangle_>-\nonumber \\ &&
	\langle \cos[2\theta(0,t_1)]\rangle_>
	\langle\partial_x\varphi(0,t_2)\rangle_>
	-\langle\partial_x\varphi(0,t_2)\rangle_>
	\langle\cos[2\theta(0,t_1)]\rangle_>
	\bigg\}{\bf  J}_y\cdot{\bf  J}_z
	\nonumber \\ &&
	-\frac{i}{\pi a}\int\! dt_1 dt_2
	\bigg\{
	\langle \sin[2\theta(0,t_1)]\partial_x\varphi(0,t_2)
	-\partial_x\varphi(0,t_2)\sin[2\theta(0,t_1)]
	\rangle_>-\nonumber \\ &&
	\langle \sin[2\theta(0,t_1)]\rangle_>
	\langle\partial_x\varphi(0,t_2)\rangle_>
	+\langle\partial_x\varphi(0,t_2)\rangle_>
	\langle\sin[2\theta(0,t_1)]\rangle_>
	\bigg\}\left({\bf  J}_z\times{\bf  J}_x\right)\cdot{\bf  S}
	\nonumber \\ &&
	+\frac{1}{\pi a}\int\! dt_1 dt_2
	\bigg\{
	\langle \sin[2\theta(0,t_1)]\partial_x\varphi(0,t_2)
	+\partial_x\varphi(0,t_2)\sin[2\theta(0,t_1)]
	\rangle_>-\nonumber \\ &&
	\langle \sin[2\theta(0,t_1)]\rangle_>
	\langle\partial_x\varphi(0,t_2)\rangle_>
	-\langle\partial_x\varphi(0,t_2)\rangle_>
	\langle\sin[2\theta(0,t_1)]\rangle_>
	\bigg\}{\bf  J}_x\cdot{\bf  J}_z.
\end{eqnarray}
The terms proportional to ${\bf  J}_x\cdot{\bf  J}_z$ and ${\bf  J}_y\cdot{\bf  J}_z$ will give rise to new coupling terms $\cos[2\theta]\partial_x\varphi$ and $\sin[2\theta]\partial_x\varphi$ which are irrelevant near $g=1$ and can be dropped. The other terms can be calculated using the commutation relations in Eq.~(\ref{app_eq:comm_with_derphi}). This leads to 
\begin{eqnarray}
	\langle \left\{S_{B},S_{F}\right\}\rangle_>\!-\!
	\left\{\langle S_{B}\rangle_>,\langle S_{F}\rangle_>\right\}
	&=& \frac{2}{\pi a}\int\! dt_1 dt_2
	\bigg\{\sum_{q<\Lambda}\frac{2\pi}{V}\cos[qu_F(t_1-t_2)]
	-\sum_{q<\Lambda '}\frac{2\pi}{V}\cos[qu_F(t_1-t_2)]\bigg\}
	\times \nonumber \\ &&
	\sin[2\theta_<(0,t_1)]e^{-g\sum_{\Lambda<q<\Lambda'}
	\frac{\pi}{qV}}
	\left({\bf  J}_y\times{\bf  J}_z\right)\cdot{\bf  S}
	\nonumber \\ &&
	-\frac{2}{\pi a}\int\! dt_1 dt_2
	\bigg\{\sum_{q<\Lambda}\frac{2\pi}{V}\cos[qu_F(t_1-t_2)]
	-\sum_{q<\Lambda '}\frac{2\pi}{V}\cos[qu_F(t_1-t_2)]\bigg\}
	\times \nonumber \\ &&
	\cos[2\theta_<(0,t_1)]e^{-g\sum_{\Lambda<q<\Lambda'}
	\frac{\pi}{qV}}
	\left({\bf  J}_z\times{\bf  J}_x\right)\cdot{\bf  S}.
\end{eqnarray}
\end{widetext}
Carrying out the sums difference in the parenthesis and taking the leading order in $dl$ we get the terms
\begin{eqnarray}
	&& \frac{2\Lambda dl}{\pi a}
	\int\! dt_1 dt_2 \cos[\Lambda u_F (t_1-t_2)] \times
	\nonumber \\ &&
	\bigg\{\sin[2\theta_<(0,t_1)]
	\left({\bf  J}_y\times{\bf  J}_z\right)\cdot {\bf  S}-
	\nonumber \\ &&
	\cos[2\theta_<(0,t_1)]
	\left({\bf  J}_z\times{\bf  J}_x\right)\cdot {\bf  S}
	\bigg\},
\end{eqnarray}
and after changing variables and integrating over $\Delta t$ in the range $[-\pi/2\Lambda u_F,\pi/2\Lambda u_F]$ we get the renormalization group equations for ${\bf  J}_x$ and ${\bf  J}_y$ as
\begin{eqnarray}
	\frac{d{\bf  J}_x}{dl} &=& (1-g){\bf  J}_x + 
	\frac{2g}{v_F}
	\left({\bf  J}_y \times {\bf  J}_z\right)\cdot {\bf  S},
	\nonumber \\
	\frac{d{\bf  J}_y}{dl} &=& (1-g){\bf  J}_y + 
	\frac{2g}{v_F}
	\left({\bf  J}_z \times {\bf  J}_y\right)\cdot {\bf  S}.
\end{eqnarray}

As a final comment we note that the contribution from $S_{F}^2$ will again be irrelevant, as it will be proportional to $(\partial_x\varphi)^2$.

\subsection{Poor Man's Scaling}
In order to corroborate our results around the noninteracting point $g=1$, we also look at the original electronic Hamiltonian of Eqs.~(\ref{eq:H0+HDelta}) and~(\ref{eq:H_S_electronic}) with no electron-electron interactions ($V=0$), and analyze it using Poor Man's scaling. To this end, we first write the Hamiltonian in dimensionless terms, where all energies are defined with respect to the high energy cutoff $D = \pi v_F / a$
\begin{eqnarray}
	\frac{\mathcal{H}}{D} &=&
	\sum_{\sigma} \int_{-1}^{1}\! dx x \varphi^{\dagger}_{\sigma}(x)\varphi(x) +
	\nonumber \\ &&
	\Delta '\int_{-1}^{1}\! dx_1 dx_2 f(x_1-x_2)
	\left[\varphi^{\dagger}_{\uparrow}(x_1)\varphi^{\dagger}_{\downarrow}(x_2) + 
	{\rm h.c.}
	\right] +
	\nonumber \\ &&
	\sum_{\alpha,\beta,\lambda,\lambda '}{J'}_{\alpha,\beta}
	\int_{-1}^{1}\! dx_1 dx_2
	\varphi^{\dagger}_{\lambda}(x_1)\varphi_{\lambda '}(x_2)
	\sigma^{\alpha}_{\lambda,\lambda '} S^{\beta},
\end{eqnarray}
where $J'_{\alpha,\beta} = J_{\alpha,\beta}/v_F$, $\Delta ' = \Delta/D$ and we have defined the dimensionless field operators
\begin{equation}
	\varphi_{\sigma}(x) = \sqrt{D} \psi_{\sigma}(xD),
\end{equation}
with $\psi_{\sigma}(\epsilon)$ the on-shell energy-field operator
\begin{equation}
	\psi_{\sigma}(\epsilon) = \sqrt{\frac{D}{2}}\sum_{k}c_{\sigma,k}
								\delta(\epsilon-\sigma\epsilon_k).
\end{equation}
The function $f(z)$ is given by
\begin{equation}
	f(z) = \frac{D}{v_F}\int\! dx
	\Theta\left(|x|-\frac{L}{2}\right)
	e^{i\Phi(x)+i\frac{D}{v_F}zx},
\end{equation}
and we emphasize it scales with $D$.

The first step is to divide the energy band into low-energy $|x| < 1-dl$ and high-energy $1-dl < |x| \leq 1$ modes, and integrate out the fast energy modes by perturbation theory. To leading order, the diagrams contributing to $\Delta$ and to the different $J$'s term do not mix, and we are left with the need to evaluate a single-type of contribution
\begin{eqnarray}
	V_{\rm eff} &=&
	-\sum_{\{\lambda_i\},\{\alpha_i\},\{\beta_i\}}
	J_{\alpha_1,\beta_2}J_{\alpha_2,\beta_2}
	\sigma^{\alpha_1}_{\lambda_1,\lambda_2}
	\sigma^{\alpha_2}_{\lambda_3,\lambda_4}
	S^{\beta_1}S^{\beta_2}
	\times \nonumber \\ &&
	\int_{-1+dl}^{1-dl}dx_{1,<}dx_{2,<}
	\varphi^{\dagger}_{\lambda_1}(x_{1,<})
	\varphi_{\lambda_4}(x_{2,<}) \times	
	\nonumber \\ &&
	\int_{1-dl}^{1}\! dx_{1,>}dx_{2,>}
	\langle\varphi_{\lambda_2}(x_>)
	\varphi^{\dagger}_{\lambda_3}(x_>)
	\rangle
\end{eqnarray}
and its corresponding contributions from the modes in $(-1,-1+dl)$. Again using the identity in Eq.~(\ref{app_eq:vec_iden}), we can carry out the multiplications and arrive at
\begin{eqnarray}
	V_{\rm eff} &=& 2dl
	\sum_{\{\alpha_i\},\{\beta_i\},\lambda_1\lambda_2}
	\epsilon_{\alpha_1,\alpha_2,\alpha_3}
	\epsilon_{\beta_1,\beta_2,\beta_3} \times \nonumber \\ &&
	{J'}_{\alpha_1,\beta_1}{J'}_{\alpha_2,\beta_2}
	\sigma^{\alpha_3}_{\lambda_1,\lambda_2}S^{\beta_3}
	\times \nonumber \\ &&
	\int_{-1+dl}^{1-dl}\!dx_{1,<}dx_{2,<}
	\varphi^{\dagger}_{\lambda_1}(x_{1,<})
	\varphi_{\lambda_1}(x_{2,<}),
\end{eqnarray}
where we have omitted constant terms and terms contributing to a scattering potential, which are irrelevant. The above expression can be written in a more concise form if we identify, similar to the bosonic case ${\bf  J}'_{\alpha} = \sum_{\beta}{J'}_{\alpha,\beta}\hat{\beta}$. We then write the effective Hamiltonian
\begin{eqnarray}
	\frac{\mathcal{H}'}{D} &=&
	\sum_{\sigma} \int_{-1+dl}^{1-dl}\! dx x 
	\varphi^{\dagger}_{\sigma}(x)\varphi(x) +
	\nonumber \\ &&
	\Delta '\int_{-1+dl}^{1-dl}\! dx_1 dx_2 f(x_1-x_2)
	\left[\varphi^{\dagger}_{\uparrow}(x_1)\varphi^{\dagger}_{\downarrow}(x_2) + 
	{\rm h.c.}
	\right] +
	\nonumber \\ &&
	\sum_{\{\alpha_i\},\lambda,\lambda '}
	\left[{\bf  J}'_{\alpha_1}+2dl\epsilon_{\alpha_1,\alpha_2,\alpha_3}
	\left({\bf  J}'_{\alpha_2}\times {\bf  J}'_{\alpha_3}\right)\right]
	\cdot {\bf  S} \nonumber \\ && \times
	\int_{-1+dl}^{1-dl}
	\! dx_1 dx_2 \varphi^{\dagger}_{\lambda}(x_1)\varphi_{\lambda '}(x_2)
	\sigma^{\alpha_1}_{\lambda,\lambda '}.
\end{eqnarray}

Finally, we rescale by $dx \to (1-dl)^{1/2} dx$, and write $\mathcal{H}'$ in terms of $D' = (1-dl)D$, to have
\begin{eqnarray}
	\frac{\mathcal{H}'}{D'} &=&
	\sum_{\sigma} \int_{-1}^{1}\! dx x 
	\varphi^{\dagger}_{\sigma}(x)\varphi(x) + \Delta '(1+dl)
	\times
	\nonumber \\ &&
	\int_{-1}^{1}\! dx_1 dx_2 f(x_1-x_2)
	\left[\varphi^{\dagger}_{\uparrow}(x_1)\varphi^{\dagger}_{\downarrow}(x_2) + 
	{\rm h.c.}
	\right] +
	\nonumber \\ &&
	\sum_{\{\alpha_i\},\lambda,\lambda '}
	\left[{\bf  J}'_{\alpha_1}+2dl\epsilon_{\alpha_1,\alpha_2,\alpha_3}
	\left({\bf  J}'_{\alpha_2}\times {\bf  J}'_{\alpha_3}\right)\right]
	\cdot {\bf  S} \nonumber \\ && \times
	\int_{-1}^{1}
	\! dx_1 dx_2 \varphi^{\dagger}_{\lambda}(x_1)\varphi_{\lambda '}(x_2)
	\sigma^{\alpha_1}_{\lambda,\lambda '},
\end{eqnarray}
where we took care to scale $f(z)$ with $(1+dl)$ as well, as it is linearly dependent on $D$. We therefore arrive at the following renormalization group equations
\begin{eqnarray}
	\frac{d\Delta}{dl} &=& \Delta,
	\nonumber \\
	\frac{d{\bf J}_i}{dl} &=& \frac{1}{v_F}\sum_{j,k}
	\epsilon_{i,j,k}{\bf  J}_j\times{\bf J}_k,
	\label{eqs_RG_Poors}
\end{eqnarray}
with the dimensions restored. These equations are similar in form to the ones derived for the bosonic Hamiltonian.

\subsection{Analysis of the renormalization group flow at $g=1$}
The set of renormalization group equations in Eq.~(\ref{eqs_RG_Poors}) does not have a closed form solution. However, some insights can be derived as to the general behavior of the coupling constants. First, as the equations satisfy
\begin{equation}
	\frac{d}{dl}\left({\bf  J}_i
	\cdot{\bf  J}_j\right) = 0,
\end{equation}
we identify the three constants of motions $C_{12}$,$C_{13}$ and $C_{23}$ as
\begin{equation}
	C_{ij} = {\bf  J}_i\cdot{\bf  J}_j = 
		|{\bf J}_i||{\bf J}_j|\cos(\theta_{ij}).
\end{equation}
Assuming that $C_{ij} \neq 0$, which represent a fine-tuned case, we note that one cannot have all $|{\bf  J}_i|$ flow to the weak-coupling fixed point of zero, as the cosine is bounded.

The differences between the magnitudes of the exchange couplings, $\delta_{ij} = {\bf J}_i^2-{\bf J}_j^2$, are also constant of motions. This means that one of them cannot increase independently, without the others growing as well. Therefore, we can divide the discussion into two separate cases:
\begin{enumerate}

\item All $|{\bf  J}_i|$ flow to infinity, while all $\cos(\theta_{ij})$ flow to zero. This is the strong-coupling isotropic fixed point.

\item The couplings flow to some intermediate fixed point where ${\bf  J}_i \parallel {\bf  J}_j$ for all $i,j$. These fixed points form a $5$-dimensional manifold within the $9$-dimensional space of coupling parameters. Such intermediate fixed point, however, is not stable. To see that we write ${\bf  J}_i = c_i {\bf  J}_0$ at the fixed point, and perturb by ${\bf  J}_z \to {\bf  J}_z+{\bf  \delta J}_z$ with ${\bf  \delta J}_z \perp {\bf  J}_0$. Linearizing the flow equations we get that about ${\bf  \delta J}_z = 0$ the perturbation follows
\begin{eqnarray}
	\frac{d{\bf  \delta J}_z}{dl} &=& 0, \nonumber \\
	\frac{d^{2}{\bf  \delta J}_z}{dl^2} &=& \frac{c_x^2+c_y^2}{v_F^2}
	|{\bf  J}_0|^2
		{\bf  \delta J}_z,
\end{eqnarray}
which leads away from the fixed point.

\end{enumerate}
Therefore, away from the fine-tuned points where $C_{ij} = 0$, the system flows to the strong coupling isotropic fixed point.

The fine-tuned points are not entirely uninteresting, as they include, for example, the isotropic Kondo model for which ${\bf  J}_i \parallel \hat{x}_i$ and all the couplings vectors are orthogonal to one another. We know that for large enough ferromagnetic $\tilde{J}_z<0$, the Kondo model flows to a fixed point with ${\bf  J}_x = {\bf  J}_y = 0$. However for weak ferromagnetic and for anti-ferromagnetic couplings the Kondo model flows to the strong coupling limit, even though it still lies in the fine-tuned manifold of $C_{ij} = 0$.

\subsection{Values of $J$ for Fig.~\ref{fig:RG_flow_examp}}
The plots in Fig.~\ref{fig:RG_flow_examp} were generated by numerically integrating the flow equations. For the continuous lines the bare initial values are
\begin{eqnarray}
	\frac{J}{v_F} &=&
		\left(\begin{array}{ccc}
		15 & 11.2 & -19.4 \\
		-0.127 & 0.075 & 0.04 \\
		-0.123 & -0.086 & 0.122
		\end{array}\right) \times 10^{-2},
\end{eqnarray}
and for the dashed lines the values are
\begin{eqnarray}
	\frac{J}{v_F} &=&
		\left(\begin{array}{ccc}
		0.15 & 0.112 & -0.194 \\
		-0.127 & 0.075 & 0.04 \\
		-0.123 & -0.086 & 0.122
		\end{array}\right).
\end{eqnarray}

\section{Electronic Solution about the Kondo fixed point}
\label{app:Kondo}
In this section we will present in further detail the procedure for the solution of the setup about the Kondo fixed point. We shall examine the Hamiltonian of a QSHI edge connecting two superconductors, with an effective potential at the origin given by Eq.~(\ref{eq:Kondo_potential}). The full Hamiltonian is then $\mathcal{H} = \mathcal{H}_0 + \mathcal{H}_\Delta + V'$ with $\mathcal{H}_0$ and $\mathcal{H}_\Delta$ given in Eq.~(\ref{eq:H0+HDelta}).

As $V'$ represent weak perturbations, the low-energy physics in this regime is dominated by the Andreev bound states. We shall project onto the finite subspace spanned by these states, and diagonalize the Hamiltonian there. To this end, we first solve the equations for the Andreev bound states without $V'$, writing the operator for an Andreev bound state with energy $\epsilon_n$ and spin $\sigma$ as
\begin{eqnarray}
	\Gamma^{\dagger}_{\sigma,\epsilon_n} &=& A_{\epsilon_n}
	\int_{-L/2}^{L/2}\!dx \bigg[\alpha_{\sigma,\epsilon_n}
	e^{i\sigma\frac{\epsilon_n x}{v_F}}
	\psi^{\dagger}_{\sigma}(x) + 
	\nonumber \\ &&
	\sigma
	e^{-i\sigma\frac{\epsilon_n x}{v_F}}
	\psi_{\bar{\sigma}}(x)
	\bigg],
\end{eqnarray}
with
\begin{eqnarray}
	\alpha_{\sigma,\epsilon_n} &=& \left[\frac{\epsilon_n}{\Delta}
	+i\sigma
	\sqrt{1-\left(\frac{\epsilon_n}{\Delta}\right)^2}\right]
	e^{-i\sigma\frac{\epsilon_n L}{v_F}},
	\nonumber \\
	A^2_{\epsilon_n} &=&
	\frac{\sqrt{\Delta^2-\epsilon_n^2}}
	{2\left(v_F+L\sqrt{\Delta^2-\epsilon_n^2}\right)}.
\end{eqnarray}
The energies are the positive solutions of the equation
\begin{equation}
	\frac{\epsilon_{\sigma}}{\Delta} = \pm \cos
	\left(\frac{\Phi}{2}-
	\sigma\frac{\epsilon_{\sigma} L}{v_F}\right),
\end{equation}
and the number of solutions increases with $L$.

Projecting onto the low-energy subspace we have
\begin{equation}
	\mathcal{H}_0+\mathcal{H}_\Delta \simeq \sum_{\sigma,n}
	\epsilon_{\sigma,n}\left(
	\Gamma^{\dagger}_{\sigma,\epsilon_{\sigma,n}}
	\Gamma_{\sigma,
	\epsilon_{\sigma,n}}-\frac{1}{2}\right).
\end{equation}
Using the inverse relations
\begin{eqnarray}
	\psi_{\sigma}(x) &\simeq& \sum_{n} A_{\epsilon_{\sigma,n}}
	\alpha_{\sigma,\epsilon_{\sigma,n}}e^{i\sigma
	\epsilon_{\sigma,n} x/v_F}
	\Gamma_{\sigma,\epsilon_{\sigma,n}} -
	\nonumber \\ && 
	\sigma\sum_{n} A^*_{\epsilon_{\sigma,n}}
	e^{-i\sigma\epsilon_{\sigma,n} x/v_F}
	\Gamma^{\dagger}_{\bar{\sigma},\epsilon_{\sigma,n}},
\end{eqnarray}
we can project $V'$ onto the low-energy subspace and write it using the creation and annihilation operators for the bound states. The effective low-energy Hamiltonian is then a matrix of dimension $2^N$ where $N$ is the nunmber of Andreev bound states that can exist in the junction, and can be diagonalized numerically. The results of such a procedure, repeated for different values of the phase bias $\Phi$, is shown in Fig.~\ref{fig:eff_H_energies}(b) for the parameters $\lambda_{PS}/v_F = -0.3$ and
\begin{equation}
	\frac{\Delta^3}{v_F^4}
	\bar{\lambda} =
	\left(
	\begin{array}{ccc}
	0.7217 & 0.2036 & 0.3743 \\
	0.2036 & 0.4904 & -0.1 \\
	0.3743 & -0.1 & 0.5133
	\end{array}\right),
\end{equation}
with $\lambda_{\alpha,\beta} = (\bar{\lambda})_{\alpha,\beta}$.

\section{Solution of the Electronic Cut-Junction}
\label{app:elec_cut_wire}

The setup discussed in this paper is composed of an electronic edge and a spin impurity. While propositions for constructing parafermions that are based on edges of fractional quantum Hall states do not have a corresponding single-body electronic picture, we may describe the states appearing in our system in electronic terms. In this section we shall present the electronic solution of the setup in the infinite backscattering limit, and show that the $\mathbb{Z}_4$ parafermions operators are many-body operators that do not have a single-body electronic structure, in contrast to Majorana bound states at the end of topological superconducting wires, for example.

Our starting point is a Hamiltonian describing non-interacting electrons in a quantum spin Hall edge. At $x=\pm L/2$ the electrons are coupled to superconducting leads with gap $\Delta$ and phases $\Phi_{R,L}$. At the origin the electrons are coupled to a backscattering magnetic impurity, and the Hamiltonian is given, in Bogoliubov de-Gennes representation, by
\begin{equation}
	\mathcal{H} = 
	\left(\begin{array}{cc}
	v_F p \sigma_z + J f(x)\sigma_x S^z &
	\Delta(x) \\
	\Delta^*(x) & -v_F p \sigma_z
	+ J f(x) \sigma_x S^z
	\end{array}\right),
\end{equation}
where the matrix is in particle-hole space, $\sigma_j$ are Pauli matrices acting in the particle spin space and $S^z$ acts on the impurity spin. Here $\Delta(x)$ is the proximity induced superconducting gap acting outside the junction, given by $\Delta(x)=\Delta \exp(i\Phi_L)$ for $x<-L/2$, $\Delta(x) = \Delta \exp(i\Phi_R)$ for $x>L/2$ and $\Delta(x)=0$ otherwise. The function $f(x)$ is strongly peaked about the origin, such that it is zero for $|x|>a/2$ for some small distance $a$.

An incoming electron from the right with momentum $p$ will penetrate the barrier with amplitude
\begin{equation}
	T(p)\sim e^{-2a\sqrt{(J/v_F)^2-p^2}}.
\end{equation}
Aiming at the infinitely strong backscattering limit, we now take the limit $J\to\infty$, such that $Ja/v_F\to\infty$ as well. In that limit the coupling to the spin-impurity induces a perfect back-scattering, where the phase-shift depends on the impurity-spin itself. The particle-states inside the junction, to the left and right of the impurity, can be written as
\begin{eqnarray}
	\psi_{p,L,s_z}(x) &=& 
	\left(\begin{array}{c}
		e^{ipx} \\ -iS^z e^{-ipx}
	\end{array}\right)
	\theta\left(x+\frac{L}{2}\right)
	\theta(-\frac{a}{2}-x),
	\nonumber \\
	\psi_{p,R,s_z}(x) &=& 
	\left(\begin{array}{c}
		iS^ze^{ipx} \\ e^{-ipx}
	\end{array}\right) 
	\theta\left(x-\frac{a}{2}\right)
	\theta(\frac{L}{2}-x).
\end{eqnarray}
Note that here we have chosen $S^z$ to have eigenvalues of $s_z = \pm 1$ (and not half that.) We then match the boundary conditions on the edges of the superconducting regions $x=\pm L/2$ and derive the full single-particle states. The many-body ground state for each impurity spin orientation $|{\rm gs}_{s_z}\rangle$ is given by filling all the states with negative energy.

However this ground state is not unique, as the sub-gap energy spectrum in each side is described by the solutions to the equation
\begin{equation}
	\frac{\epsilon}{\Delta} = \pm
			\sin\left(\frac{\epsilon L}{2v_F}\right),
\end{equation}
which always support a zero-energy state. The zero-energy solutions on each side are Majorana bound-states $\gamma_{L,R}$, whose wave function is dependent on the orientation of the impurity spin. Inside the junction the wave functions for these states are given by
\begin{eqnarray}
	\psi_{\gamma_L}(x) &=& e^{i\frac{\pi}{4}(1-S^z)}\!
	\left(\!\begin{array}{c}
	e^{\frac{i}{2}\Phi_L} \\ -iS^ze^{\frac{i}{2}\Phi_L} \\
	ie^{-\frac{i}{2}\Phi_L} \\ -S^ze^{-\frac{i}{2}\Phi_L}
	\end{array}\!
	\right)\!\!
	\theta\left(x+\frac{L}{2}\right)
	\theta(-\frac{a}{2}-x),
	\nonumber \\ 
	\psi_{\gamma_R}(x) &=& e^{i\frac{\pi}{4}(1-S^z)}\!
	\left(\!\begin{array}{c}
	iS^z e^{\frac{i}{2}\Phi_R} \\ e^{\frac{i}{2}\Phi_R} \\
	S^z e^{-\frac{i}{2}\Phi_R} \\ ie^{-\frac{i}{2}\Phi_R}
	\end{array}\!
	\right)\!\!
	\theta\left(x-\frac{a}{2}\right)
	\theta(\frac{L}{2}-x),
\end{eqnarray}
and outside the junction they decay exponentially over the superconductor coherence length $\xi=v_F/\Delta$. Here the wave function is given in Nambu representation $\Psi = (\psi^{\dagger}_{\uparrow} ,\; \psi^{\dagger}_{\downarrow} ,\; \psi_{\downarrow} ,\; -\psi_{\uparrow})^T$ Each pair of Majorana bound-states creates one non-local fermionic state $i\gamma_L\gamma_R = 2\Gamma_0^{\dagger}\Gamma_0-1$ which can be either occupied or empty. Therefore we can, in that basis, see explicitly the four-fold degenerate ground-state subspace, where the $2$-fold degeneracy due to the spin orientation is multiplied by the $2$-fold degeneracy described by the occupancy of the fermionic state, and we denote them by $|{\rm gs}_{n,s_z}\rangle$ with $n=0,1$.

It is worthwhile to consider the behavior of these states under time reversal symmetry. To this end, we start by noticing that
\begin{equation}
	T\gamma_{L,R}(\Phi_{L,R})T^{-1} =
	-S_z \gamma_{L,R}(-\Phi_{L,R}),
\end{equation}
where we wrote explicitly the dependence of the Majorana wave function on the phase of the adjacent superconducting lead. This implies $T\Gamma T^{-1} = -S_z\Gamma^{\dagger}$. In addition to that, $TS_zT^{-1} = -S_z$, which leads to the observation that time reversal changes both the occupancy of the fermionic state and the projection of the impurity spin, with a sign that depends on $S_z$.

Even though there is a single-body description of the Majorana bound states, they cannot connect between all the states within the ground state manifold. They are bound to a specific value of $S^z$, and therefore can relate $|{\rm gs}_{0,s_z}\rangle$ with $|{\rm gs}_{1,s_z}\rangle$ but not with $|{\rm gs}_{n,-s_z}\rangle.$ The parafermionic operators cannot be described as a linear combination of single-body operators. To see this, we note that due to the presence of the superconductor, the different states in the ground states manifold are condensates, and their wave function explicitly depends on $S^z$. Relating ground states with different impurity spin orientation requires taking care of the scattering phase shifts of all states in the condensate. Based on the bosonic picture, we may identify the fermionic representation of the operator $\hat{F}$ that transfer the system within the degenerate ground-state manifold as given in Eq.~(\ref{eq:F_fermionic}). This operator is indeed a many-body operator, that changes the relative phase of the reflected electrons throughout the entire condensate.

%\bibliographystyle{h-physrev}
%\bibliography{mybib}

\end{document}